# On a revised concept of an event that allows linking nanodosimetry and microdosimetry in nanometric sites with macroscopic dosimetry

Hans Rabus*, Leo Thomas

*Physikalisch-Technische Bundesanstalt (PTB), 10587*

*Berlin, Germany*

*email: hans.rabus@ptb.de

## Abstract

Microdosimetry and nanodosimetry study the track structure of charged particles, i.e., the stochastics of radiation interaction on the microscopic scale. The (different) concepts developed in micro- and nanodosimetry are matched to the experimental approaches in the two fields. This work reviews the concepts of an event used in the two fields and analyzes how microdosimetric and nanodosimetric single event distributions could theoretically be derived from probability distributions related to interactions of the primary particle which produce secondary electrons. It is shown that the corresponding mathematical expressions of conditional ionization cluster size distributions are alike those for the single event frequency distribution of energy imparted. This is particularly true when all tracks are considered which intersect the volume in which interactions of the primary particle can result in energy deposits or ionizations in the site. This volume, named beam volume of the site, cannot be determined explicitly. Therefore, track structure simulations of proton with energies between 1 MeV and 100 MeV are used to study how the occurrence of events depends on site size, beam radius, and proton energy. For small nanometric targets, most events according to the conventional definition of microdosimetry were found to produce energy imparted below the first electronic excitation level of water and therefore appear biologically irrelevant. On the other hand, the range of impact parameters of particle tracks that contribute to energy imparted in a site appears not to depend on whether any energy deposits or only energy deposits by ionizations are considered. Since there is no longer a one-to-one correspondence between tracks passing a site and the occurrence of an event, it is proposed to use the fluence for which on average one event occurs as a substitute for single events. For protons, the product of this fluence and the site cross section or the average number of tracks necessary for an event shows an interesting dependence on site size and particle energy with asymptotic values close to unity for large sites and proton energies below 10 MeV. For a proton energy of 1 MeV, a minimum of the number of tracks is observed for sites between 5 nm and 10 nm diameter. The relative differences between the numbers of track per event on average obtained with different options of Geant4-DNA are in the order of 10 % and illustrate the need for further investigations into cross-section datasets and their uncertainties.

# 1 Introduction

Microdosimetry and nanodosimetry are fields of radiation dosimetry. Both study the track structure of charged particles, i.e., the stochastics of radiation interaction on the microscopic scale (Booz et al., 1983; Braby et al., 2023).

The track structure is related to the radiobiological effectiveness (RBE) of different radiation qualities, and microdosimetric measurement quantities were explored as measurands related to RBE (Menzel et al., 1990; Allisy et al., 1993; Loncol et al., 1994; Hawkins, 1998; Parisi et al., 2022). Similarly, a number of approaches have been proposed to establish a link between nanodosimetry and RBE (Schulte et al., 2001, 2008; Grosswendt, 2005, 2006; Garty et al., 2006, 2010; Besserer and Schneider, 2015; Conte et al., 2017, 2018, 2023; Schneider et al., 2019).

Within the framework of the European project "Biologically weighted quantities in radiation therapy (BioQuaRT)" (Rabus et al., 2014), a generic multi-scale approach was proposed for the RBE of ion beams that encompasses both microdosimetry and nanodosimetry (Palmans et al., 2015). The BioQuaRT approach pertained to the relation between the distribution of energy imparted in a larger microdosimetric site and the ionization cluster size (ICS) distributions in nanometric sites contained in it. An alternative approach is to study the connection of the distributions of energy imparted and ICS in the same site (Amols et al., 1990; Selva et al., 2022a, 2023).

The (different) concepts of measurement quantities developed in micro- and nanodosimetry are matched to the experimental approaches in the two fields.

Microdosimetry characterizes particle track structure by the frequency distribution of the energy imparted by ionizing particles in a specified target called the site (Rossi, 1960; Rossi and Zaider, 1996; Lindborg and Waker, 2017). The site dimensions in tissue are typically in the order of several micrometers, that is, of the dimensions of cell nuclei or chromosome domains within them. The energy imparted in such a site is a stochastic quantity resulting from one or several events, which are defined by passages of primary particle tracks through or nearby the site that result in energy deposits in the site (Fig. 1). The multi-event frequency distribution can be related to the conventional dosimetric quantity absorbed dose and it can be derived from the single-event frequency distribution of energy imparted by single tracks based on the statistical independence of different events (Rossi and Zaider, 1996; Lindborg and Waker, 2017).

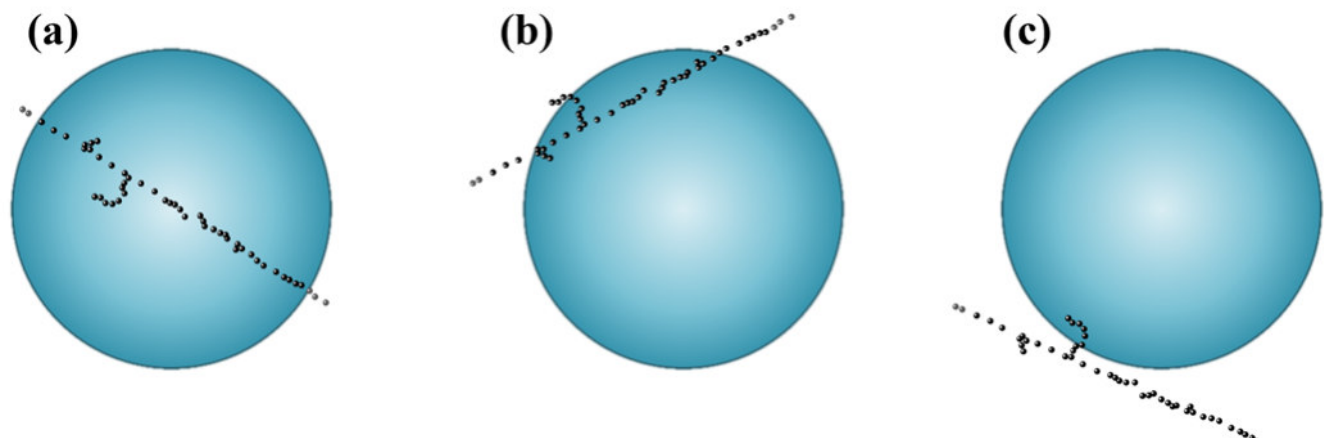

**Fig. 1:** Illustrations of microdosimetric single events in spherical sites (blue spheres). The black dots represent the energy transfer points in the charged particle tracks traversing the site in (a) and (b). In (c) the energy deposits are only from delta electrons produced by the primary particle outside the site.

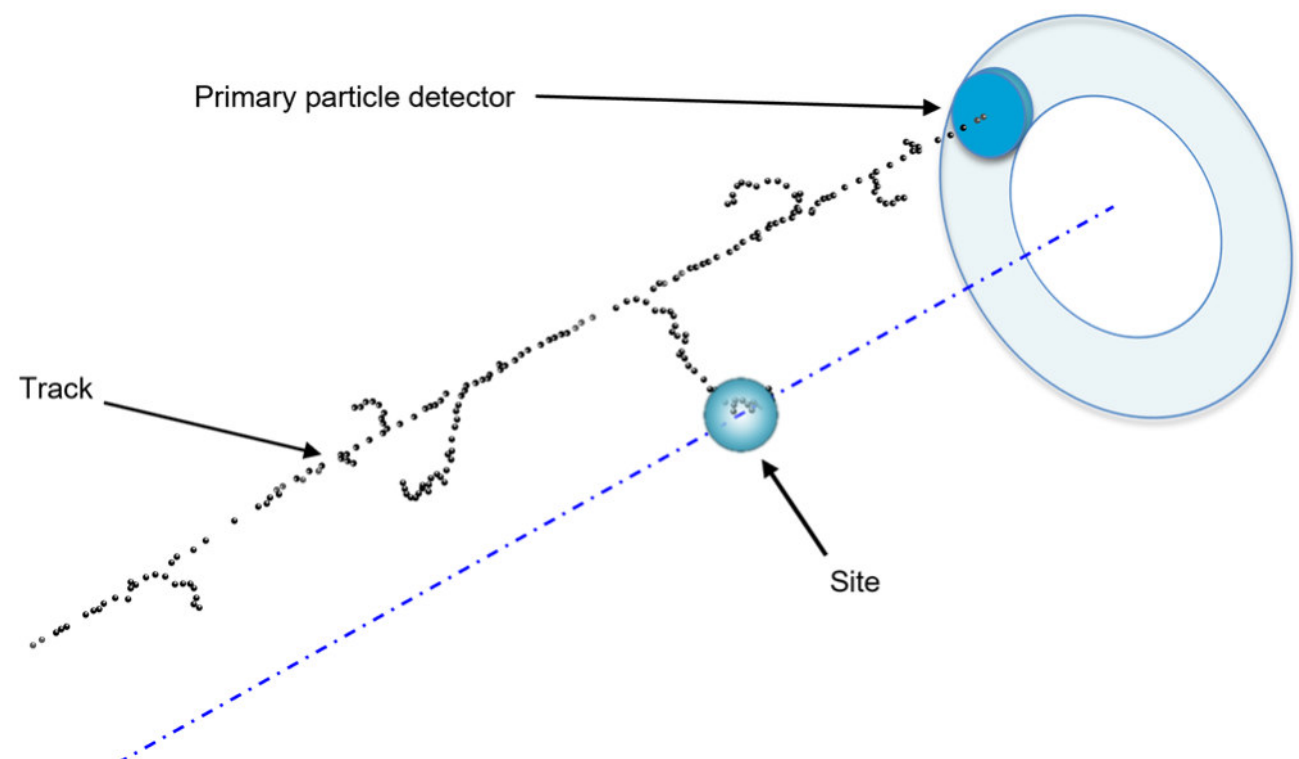

**Fig. 2:** Illustration of a nanodosimetric event in a spherical site passed by a track at a given impact parameter. The black dots represent energy transfer points with ionizing interactions. The ionizations in the site are scored in coincidence with the detection of a primary particle passage. The annulus represents the area covered by primary particles detectors corresponding to the same impact parameter.

Nanodosimetry was developed in response to evidence that clusters of lesions on the DNA molecule within few nanometers are the starting point of biological radiation effects (Goodhead, 1989, 1994, 2006). In conventional nanodosimetry, the frequency distribution of the number of ionizations in a site is the quantity of interest, traditionally referred to as ICS distribution, which is also used in this work instead of the recently proposed term "ionization detail" (Rucinski et al., 2021). The considered site dimensions are generally in the few to few tens of nanometer range (Grosswendt, 2002; Bantsar et al., 2018; Pietrzak, 2019; Rabus, 2020). ICS distributions are considered for a specific impact parameter of a particle track with respect to the center point of the target volume (Fig. 2). Therefore, these distributions are conditional on the knowledge that the passage of a track occurred. In consequence, the ICS distribution includes a non-zero frequency for ICS zero, in contrast to the microdosimetric single-event distribution of energy imparted.

Nanodosimetric detectors measure the ions produced in a target volume by a passing track (Bantsar et al., 2018). Microdosimetric detectors also measure electrical charge. For conventional gas-based microdosimeters this is the number of electrons generated by gas amplification of the ionizations produced by the particle track in the detector's sensitive volume. This charge is then converted to energy imparted by a calibration factor derived from the energy position of full-absorption edges of ionizing particles (Bianchi et al., 2021).

Conceptionally, energy imparted in a detector can also be determined from the number of ionizations produced in it by radiation interactions, namely by multiplication with the mean energy imparted per ionization. This approach faces challenges when site sizes in the nanometer range are considered (Amols et al., 1990). Ongoing endeavors to develop microdosimetric detectors capable of measuring distributions for simulated site sizes in the range below 100 nm (Bortot et al., 2017, 2022; Mazzucconi et al., 2020a, 2020b; Bianchi et al., 2020) have prompted a reconsideration of this issue in recent work (Selva et al., 2022b; Lillhök et al., 2022; Lindborg and Rabus, 2023; Bianchi et al., 2024).

While nanodosimetry was originally developed to provide a better measure of radiation quality, several authors have explored the use of nanodosimetric track structure information for predicting cell survival or radiobiological effectiveness (Alexander et al., 2015a; Besserer and Schneider, 2015; Conte et al., 2017, 2024; Casiraghi and Schulte, 2015; Schneider et al., 2019, 2020; Selva et al., 2019; Mietelska et al., 2024). Recently, a theoretical framework has been proposed for nanodosimetry-based treatment planning in hadron therapy which builds on the combination of particle fluence and nanodosimetric ICS distributions (Faddegon et al., 2023).

Motivated by these developments, this paper explores a revised concept of a nanodosimetric event that enables relating nanodosimetric quantities to macroscopic dosimetric concepts such as fluence and absorbed dose. Simulations of proton tracks are then used to explore the question of how the fluence corresponding to the occurrence of one event on average depends on radiation quality. In addition, it is explored how the results depend on the choice of interaction cross-sections in the simulation code used.

## 2 Materials and Methods

### *2.1 The event concept in microdosimetry and nanodosimetry*

The difference between microdosimetry and nanodosimetry with respect to the concept of an event is evident from Figs. 1 and 2. In a microdosimetric (single) event, the measured signal generally contains no information on the geometrical relation between the track and the site[1]. Therefore, the microdosimetric concept of lineal energy is defined as the ratio of the energy imparted by an event to the mean chord length through the target volume.

In a conventional nanodosimetric measurement, the geometrical relation between track and site is known[2]. In addition, it is generally unlikely that several, statistically independent events interact with the same site owing to its small size (Ngcezu and Rabus, 2021). By performing measurements at different impact parameter or by using position-sensitive primary particle detectors, it is possible to investigate 'broad field' irradiations (Hilgers and Rabus, 2020; Hilgers et al., 2022). However, it is not straightforward to link ICS distributions and macroscopic fluence.

The following section outlines a theoretical consideration on the relation between a charged particle track and a microscopic or nanoscopic site which results in a contribution to energy imparted or a certain number of ionizations in the site. This provides a model for the potential use of an analog to the microdosimetric concept of an event also in nanodosimetry. A detailed mathematical argument can be found in the Appendix, whereas here only the main ideas are presented.

[1] The exception is the novel type of microdosimeter presented by (Mazzucconi et al., 2020b), in which measurements are performed with defined impact parameter.

[2] For completeness, it should be noted that nanodosimetric measurements without a trigger detector have also been explored (Pietrzak, 2019).

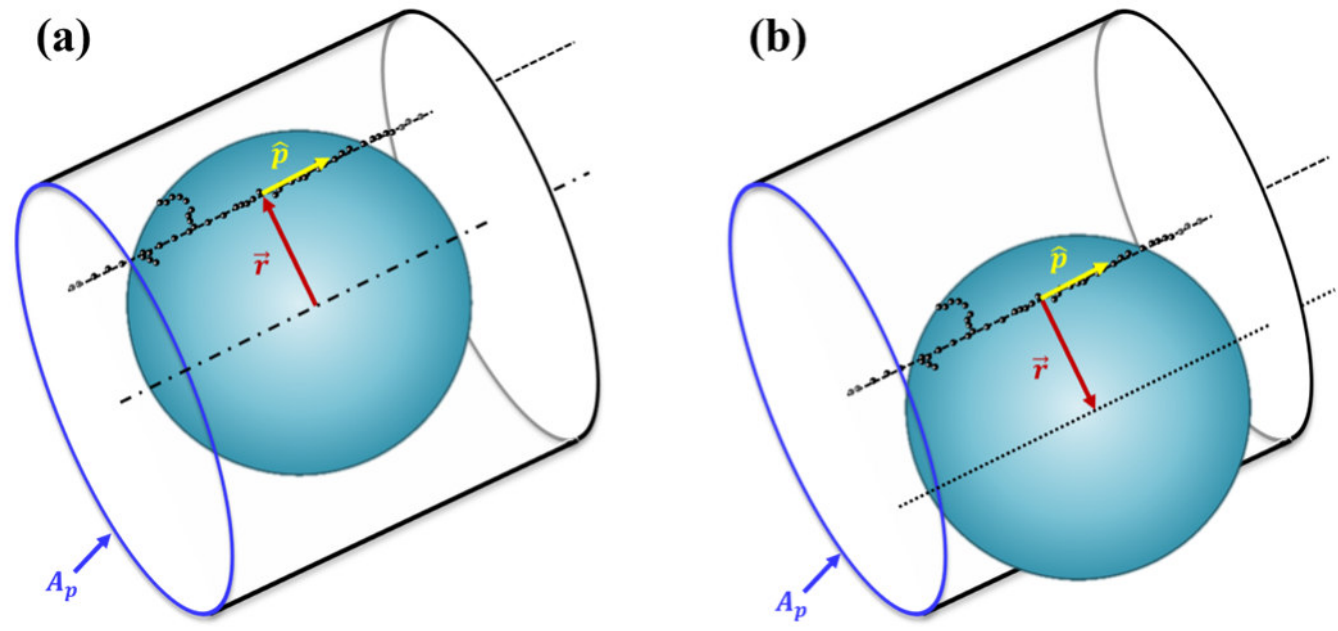

**Fig. 3:** Illustration of the definitions of a (a) cylindrical beam volume related to the site (BVS) and (b) cylindrical scoring volume around a track (SVT). The cylinders have a cross-sectional area $A_p$ and a rotation axis given by the direction of motion of the primary particle $\hat{p}$. The site is traversed by a primary particle at an offset $\vec{r}$ from its center. In (a) $\vec{r}$ describes the location of the track relative to the site, in (b) it describes the location of the site relative to the track.

The starting point is that, by definition, a charged particle track is a collection of statistically correlated energy transfer points (ETPs). If a single charged particle track passes or traverses a site as illustrated in Fig. 1 and Fig. 2, the number of ionizations and energy imparted in the site by this track are stochastic quantities. The probability distributions of these stochastic quantities are conditional on the geometrical relation between the particle track and the site. If the primary particle travels in a straight line, the impact parameter (i.e., length of the vector $\vec{r}$ in Fig. 3 (a)) is sufficient to characterize this geometrical relation.

ETPs inside the site can be either from an interaction of the primary particle or from an interaction of an electron which is a descendant of a secondary electron produced in an interaction of the primary particle. The probability of an ETP in the site produced by such an electron decreases with increasing distance between the site center and the point where the primary particle interacted and produced the secondary electron. Therefore, there is a volume associated with the site comprising all points at which a primary particle interaction taking place and producing an electron has a non-zero probability of resulting in an ETP in the site.

In the mathematical argument in the Appendix (Supplement 1) and in the following, it is convenient to consider the smallest cylinder enveloping this volume for a given direction of motion of the primary particle (or unidirectional irradiation), as illustrated in Fig. 3 (a). For ease of reference, this cylinder is referred to as the beam volume of the site (BVS). As illustrated in Fig. 3 (a), the BVS is characterized by its cross section $A_p$ and the direction of motion $\hat{p}$ of the primary particles. By definition, its rotation axis (dot-dashed line in Fig. 3 (a)) passes through the center of the site, and the cylinder extends over a stretch of its axis in both directions from the site center.

### 2.1.1 *Single events in microdosimetry*

In Appendix A.1 it is shown that, for unidirectional irradiation, the single event probability distribution $f_1(\varepsilon)$ of energy imparted, $\varepsilon$, can be expressed as the average of the contributions to this probability from tracks intersecting the cross-section of the BVS. The corresponding expression is given in Eq. (1). As indicated on the left-hand side of the equation, $f_1(\varepsilon)$ is conditional on the parameters $Q$ and $d_s$ (omitted on the right-hand side for better legibility), where $Q$ denotes the radiation quality (type and energy of the primary particle) and $d_s$ the site diameter.

$$f_1(\varepsilon|Q,d_s) = \frac{1}{1-F_0}\frac{1}{A_p}\int\limits_{A_p} f_1^*(\varepsilon|\vec{r})\; d^2\vec{r} \tag{1}$$

In Eq. (1), $\vec{r}$ is a vector connecting the center of the site and the closest point to it on a primary particle trajectory (PPT). The length of this vector is the impact parameter. $f_1^*(\varepsilon|\vec{r})$ is the probability density distribution of energy imparted conditional on a primary particle trajectory passing the center of the target volume with a lateral offset $\vec{r}$. (The asterisk indicates that the corresponding probability has a non-zero value at zero energy imparted.) $A_p$ is the cross-section of the BVS, and $F_0$ is the average probability of no energy being deposited in the site from a PPT traversing the BVS along direction $\hat{p}$. The factor $1/A_p$ corresponds to a fluence value of one particle crossing the area $A_p$.

It should be noted that if $A_p$ was replaced in Eq. (1) by any larger area that fully contains $A_p$, the denominator $(1-F_0)$ decreased by the same factor as the area increased. This means that the $f_1(\varepsilon)$ resulting from Eq. (1) does not depend on the lateral extension of the BVS as long as the BVS covers all tracks that may contribute to the energy imparted in the site.

### *2.1.2 Single events in nanodosimetry*

Using a notation analogous to that applied for microdosimetry, Appendix A.2 (Supplement 1) shows that the ICS distribution in conventional nanodosimetry can be expressed by Eq. (2).

$$p_1(\nu|r_A;Q,d_s,A) = \frac{1}{A}\int\limits_{A} p_1^{\#}(\nu|\vec{r})\; d^2\vec{r} \tag{2}$$

That is, as the average of the contributions from tracks intersecting the area $A$ of the primary particle detector (cf. Fig. 2). Here, $r_A$ is the lateral offset between the centers of the site and the center of the area $A$, and $p_1^{\#}(\nu|\vec{r})$ is the conditional probability of $\nu$ ionizations to occur in a site passed by a track with an offset from the site's center given by $\vec{r}$. The superscript "#" indicates that in this case the probability distribution also contains an entry for no ionizations occurring. It should be noted that $p_1(\nu)$ is conditional on $Q$ and $d_s$ and, in addition, also on $r_A$ and $A$.

The corresponding ICS distribution conditional on the occurrence of at least one ionization can then be written as Eq. (3).

$$p_1(\nu|r_A;Q,d_s,A) = \frac{1}{1-P_{0,\mathrm{d}}(r_A;Q,d_s,A)}\frac{1}{A}\int\limits_{A} p_1^{\#}(\nu|\vec{r})\; d^2\vec{r} \tag{3}$$

Here, $P_{0,\mathrm{d}}$ is given by Eq. (A.16) and represents the average probability for obtaining zero ionizations in the site when a primary particle hits the primary particle detector (hence, the index d).

The ICS distribution in Eq. (3) is conditional both on the occurrence of at least one ionization in the site and on the passage of the primary particle through area $A$. However, it is also possible to consider the passage of the primary particle through the cross-sectional area $A_p$ of the BVS (Fig. 3(a)) and to define a single event ionization cluster size distribution $p_1(\nu)$ by

$$p_1(\nu|Q, d_s, A_p) = \frac{1}{1-P_0}\frac{1}{A_p}\int_{A_p} p_1^{\#}(\nu|\vec{r})\, d^2\vec{r} \tag{4}$$

with $P_0$ given by Eq. (A.19) representing the probability of zero ionizations in the site when a primary particle track traverses the BVS. The ICS distribution given in Eq. (4) is defined in full analogy to the single event distribution for energy imparted in that it contains the contributions from all particle tracks that can contribute to ionizations in the site.

### *2.1.3 Multi event distributions in microdosimetry*

In microdosimetry, multi-event distributions are derived from the single-event distributions $f_1(\varepsilon)$ based on the statistical independence of events. The resulting multi-event distribution is given by Eq. (5) and is conditional on the expected number $\bar{n}$ of events (event frequency) which depends on the fluence of the radiation field. In Eq. (5), the quantities $\varepsilon_1 \dots \varepsilon_n$ are now the energies imparted by each of the *n* events and all have nonzero values. Dirac's δ function is used in Eq. (5) instead of the conventionally used formulation with multiple convolutions.

$$\begin{aligned} f(\varepsilon|\bar{n}) = \sum_{n=1}^{\infty} \bar{n}^n e^{-\bar{n}} \int_0^{\varepsilon} f_1(\varepsilon_1) \int_{\varepsilon_1}^{\varepsilon} f_1(\varepsilon_2) \cdots \times \int_{\varepsilon_{n-1}}^{\varepsilon} f_1(\varepsilon_n) \\ \times\, \delta\left(\varepsilon - \sum_i \varepsilon_i\right) d\varepsilon_1\, d\varepsilon_2 \cdots d\varepsilon_n \end{aligned} \tag{5}$$

The summation index in Eq. (5) is the number of events interacting with the site. The event frequency $\bar{n}$ is generally estimated from the fluence $\phi$ of incident primary particles and the cross-section $A_s$ of the site,

$$\bar{n} = \phi \times A_s\,, \tag{6}$$

whereby it is implicitly assumed that $A_s$ and the area $A_p$ are the same.

This assumption can be made when the range of secondary electrons is small compared to the site size. For site sizes in the nanometer regime, the energy imparted on average by a single event may still be approximately the same as the energy transferred to secondary electrons by interactions of the primary particle in the site. However, energy imparted is defined as the sum of the energy deposits in the site by a single event (Booz et al., 1983; Braby et al., 2023).

In events with a PPT passing the site without traversing it, energy may be imparted in the site by secondary electrons only (Fig. 1(c)). This is only the case for a small number of events. On the other hand, PPTs intersecting the cross-sectional area of the site near its perimeter have a small probability of not leading to energy deposits in the site and, hence, of not representing an event. The likelihood for this to occur can be expected to increase with decreasing site size.

This is because the range of possible distances a point in the site cross-section may have from the perimeter of the cross-section decreases. Therefore, the proportion of tracks intersecting the site cross-section closer to its perimeter increases.

For sites with dimensions of few to few tens of nanometers, all points on the site cross-section may be close to its perimeter according to such a criterion. Under such conditions, using the cross-sectional area of the site in Eq. (6) becomes inadequate.

#### *2.1.4 Multi-event distributions in nanodosimetry*

The probability of different PPTs traversing the same cross-sectional area of nanometric dimensions is negligibly small for fluences corresponding to doses typically considered in therapeutic applications and radiation protection (Ngcezu and Rabus, 2021). Therefore, an event frequency defined according to Eq. (6) with the area $A_s$ replaced by the area $A$ of the primary particle detector is generally a very small number.

However, as already indicated by the annulus shown in Fig. 2, PPTs of a given impact parameter with respect to the site cover a much larger area than the primary particle detector alone. The only exception to this is a primary particle detector placed such as to detect PPTs traversing the site. For all impact parameters, the area of the annulus and, thus, the probability of a track passing through it grows proportional to the impact parameter.

For relating nanodosimetry with fluence, it therefore appears more appropriate to reconsider the concept of a nanodosimetric event as suggested above with Eq. (4). Such an approach corresponds to what was termed "broad beam" geometry (Hilgers et al., 2017) and is proposed here as the starting point for linking nanodosimetry and macroscopic fluence which results in an expression for the multi-event frequency distribution of ionization cluster size as given in Eq. (7).

$$p(\nu|\bar{n}) = \sum_{n=1}^{\infty} \bar{n}^n e^{-\bar{n}} \sum_{\nu_1,\nu_2,\ldots,\nu_n}^{\infty} p_1(\nu_1) p_1(\nu_2) \cdots \times p_1(\nu_n)\, \delta_{\nu,\sum_i \nu_i} \tag{7}$$

Here $\bar{n}$ is the event frequency, $p_1(\nu_i)$ is the probability of the $i^{\text{th}}$ event to produce an IC of $\nu_i$ in the site, and Kronecker's δ is used to avoid the clumsy notation of multiple convolutions.

#### *2.1.5 Effective beam cross-section corresponding to a single event*

With nanometric sites, the occurrence of an event is not related to the passage of the PPT through the site. Conversely, not all PPTs that pass the cross-sectional area $A_p$ lead to an event. Therefore, to relate single and multiple events it appears necessary to consider the fluence $\phi_1 = 1/A_1$ which results in an event frequency of unity. If an event is defined by the occurrence of at least one ionization in the site, then the area $A_{1,i}$ can be expected to be almost identical to the microdosimetric area $A_{1,\varepsilon}$ defined by the expectation of an event number of unity for events defined by the occurrence of an energy deposit in the site. Since for small sites energy deposits by (non-ionizing) electronic excitations may be more important than for larger sites (Lillhök et al., 2022), $A_{1,\varepsilon}$ may differ from $A_{1,i}$. On the other hand, events with only non-ionizing energy deposits in the site are not detectable with present state microdosimeters so that it may also make sense to consider only detectable microdosimetric events, in which case the relevant area is $A_{1,i}$.

It should be noted that studying the connection between microdosimetric and nanodosimetric distributions for the same site (implicitly) also implies the use of the same event concept. For instance, (Selva et al., 2022b) considered the relation between $f_1(\varepsilon|\vec{r})$ given by Eq. (A.7) and $p_1^{\#}(\nu|\vec{r})$ given by Eq. (A.13) for the case of the PPT centrally passing the site. (Using a constant $W$-value as conversion factor between the distributions of ICS and energy imparted.) In a follow-up paper (Selva et al., 2023), extended beams (larger than the site size) were considered. Thus, instead of conventional nanodosimetric quantities an average over a range of impact parameters was considered, as in Eq. (4) but with the difference that the range of impact parameters was chosen arbitrarily instead of by the more objective criterion described above.

#### *2.1.6 Target-centered versus track-centered view*

The BVS was introduced above as a theoretical concept to facilitate the mathematical arguments in the Appendix. The cross-sectional area of the BVS cannot be determined explicitly. Estimating the BVS by simulations of, for instance, pencil beams at different impact parameters would be extremely computationally inefficient. Therefore, an alternative approach as illustrated in Fig. 3(b) is used here, which considers a scoring volume around a track (SVT), comprising all possible locations of site centers within a cylinder around a stretch of the PPT. Computationally efficient scoring can then be achieved by segmentations of the volume around a PPT using virtual targets derived from the positions of the ETPs in a track (Braunroth et al., 2020; Ngcezu and Rabus, 2021) or by using the so-called associated-volume clustering approach (Lea, 1940, 1955; Kellerer, 1985; Rossi and Zaider, 1996; Thomas et al., 2024). In both cases, ETPs at all radial distances from the PPT are included.

The key rationale of this alternative approach is that the probability of an ICS formed in a site when a PPT passes the site with a given impact parameter does not depend on the choice of coordinate system. In an experiment, or a simulation aimed at replicating the experiment, it is natural to consider the target at the center and the track passing at some distance from this center as shown in Fig. 2. However, the same probability of an ICS is obtained when the PPT passes the origin, and the site has a certain offset from the axis defined by the PPT.

As first realized for nanodosimetry by (Selva et al., 2018), the expected relative frequency of tracks passing uniformly distributed within a given area relative to the site that result in an ICS of $\nu$ is therefore the same as the expected relative frequency of targets with an ICS of $\nu$ among multiple target volumes uniformly distributed over the same area relative to the PPT. For instance, the ICS distribution obtained by tracks passing the circular cross-sectional area of the BVS in Fig. 3(a) is the same as the relative frequency distribution of targets with the respective ICS when the targets are uniformly distributed across a circle of the same area which is centered on the PPT and perpendicular to it (Fig. 3(b)).

Over segments of the PPT along which the change in energy of the primary particle is negligible small, the underlying probability distribution of ICS can be assumed to be constant along this path segment. This was exploited in previous work (Rabus et al., 2020; Braunroth et al., 2020, 2021) to improve the scoring efficiency by averaging over a stretch of the PPT when determining the ICS probabilities in cylinder shells around the PPT to determine the (cylinder-)radial dependence of ICS probabilities around the PPT.

It is worth mentioning that to some extent, the concept of an 'effective track cross-section' used in (Rabus et al., 2020; Braunroth et al., 2020, 2021) implicitly contained the present proposal. This quantity can be easily interpreted as the ratio of the frequency of ionization clusters in nanometric sites to the fluence of primary particles. As can be derived from these preceding studies, the cross-sectional area $A_{1,\nu}$ corresponding to an event frequency of unity for formation of an ionization cluster of size $\nu$ depends on cluster complexity. Investigating this aspect further is beyond the scope of the present study.

A deficiency of the approach by (Rabus et al., 2020; Braunroth et al., 2020, 2021) and that developed from it in (Ngcezu and Rabus, 2021) was that only conditional ICS probabilities could be assessed. This problem can be overcome by using the concept of a SVT around the PPT if the energy loss of the primary particle is negligible small along the corresponding stretch of the PPT. Conceptionally, this SVT of a PPT can be segmented into sub-volumes $V_\nu$ defined as the union of all points in the SVT around which a sphere of the site radius contains $\nu$ ionizations. The expectation of the ratio $V_\nu/V_{\mathrm{SVT}}$ is the probability of obtaining an ICS of $\nu$. The probability of ICS zero is the corresponding ratio for the union of all points with no ionization closer than the site radius.

### *2.2 Simulations and data analysis*

Simulations of proton tracks in liquid water were performed with the GEANT4-DNA toolkit (Incerti et al., 2010a, 2010b; Bernal et al., 2015; Incerti et al., 2018). Version 10.4 patches 01 and 02 of the code were used with options 2, 4 and 6 of the G4EmDNAPhysics constructors. These physics constructors use cross-section models for electron transport in liquid water that allow a step-by-step simulation of the radiation interactions.

The same set of initial proton energies (between 1 MeV and 99 MeV) as used in earlier work (Rabus et al., 2020; Braunroth et al., 2020, 2021; Ngcezu, 2021) was also used here. Unlike these studies, electrons were followed in the entire world volume until their kinetic energy fell below a threshold of 1 eV. When this occurred, their residual energy was deposited on the spot. The transport of protons was terminated when they crossed a plane perpendicular to the initial direction of motion at 1 μm from the proton's start point. For each energy transfer point (ETP) from proton and electron interactions an n-tuple consisting of proton event ID, ETP position, type of interaction, deposited energy, energy loss of the incident particle and its energy before the collision were registered.

The data analysis was performed using FORTRAN code. Each of the 50,000 simulated proton tracks was processed separately to obtain energy imparted and ICS in spherical targets of different sizes. A variant of the associated volume clustering (AVC) algorithm (Kellerer, 1985; Rossi and Zaider, 1996) was used to identify clusters of ETPs.

In a first preprocessing step, ETPs from electron interactions outside the plane parallel slab of 1 μm thickness (defined by the proton track segment) were moved by integer multiples of 1 μm such that the shifted points were within the slab. This avoids the rather complex procedure proposed in (Braunroth et al., 2020) to compensate for the secondary electron disequilibrium due to the lack of interactions from electrons originating in proton interactions before or behind the track segment considered in the simulation.

In a second preprocessing step the ETPs of a track were sorted with respect to their position along the direction of motion of the proton track. This allows restricting the search for cluster partners to those ETPs that have an offset from the sampled target center in the direction of the proton momentum of maximum the site radius.

The AVC algorithm was used such that all ETPs of a track were processed consecutively. For each of them, a potential site center was uniformly random sampled within a sphere having the radius of the site. The ETPs in the interval of indices established in the second preprocessing step that had a distance less than or equal to the site radius were determined, and their energy deposits were added to obtain the energy imparted. The AVC algorithm was applied repetitively (with $n_r$ repetitions) to average over possible site placements.

The number of ETPs related to an ionization gave the ICS, where ionizations of the oxygen 1s orbital were counted twice owing to predominant de-excitation by the Auger-Meitner process. Furthermore, excitations to molecular orbitals at energies above the first ionization threshold were assumed to autoionize with 100 % probability. Energy imparted and ICS were registered with a weight equal to the inverse of the number of ETPs in the site to account for the fact that the site center is in the intersection volume of this number of spheres around the ETPs.

The analysis was performed for site diameters $d_s$ of 1 nm, 2 nm, 3 nm, 5 nm, 10 nm, 20 nm, 30 nm, 50 nm, and 100 nm. The beam diameters were between 1 nm and 50 μm, using intermediate values at two, three, and five times the start of each decade. The beam radius corresponds to the maximum impact parameter of the proton with respect to the center of the site.

From the quantities scored per track, the frequency distributions of ICS and energy imparted (and the derived quantities lineal and specific energy) were determined along with the bin-wise standard deviation as an estimate of the uncertainty. The frequency distributions of microdosimetric quantities were determined conditional on the ICS in the site and conditional on the site being in a cylindrical SVT (Fig. 3(b)).

$$\bar{f}_{1\nu}\left(\varepsilon_j \middle| d_s, r_b\right) = \frac{N_{\nu,j}}{N_\nu \, \Delta\varepsilon_j} \qquad\qquad N_\nu = \sum_j N_{\nu,j} \tag{8}$$

In Eq. (8), $N_{\nu,j}$ is the number of sites of diameter $d_s$ with an ICS of $\nu$ and an energy imparted between $\varepsilon_j - \Delta\varepsilon_j/2$ and $\varepsilon_j + \Delta\varepsilon_j/2$ in a cylinder of radius $r_b$ around the proton trajectory. $\varepsilon_j$ and $\Delta\varepsilon_j$ are the center and width of the $j^{\text{th}}$ energy bin. $N_\nu$ is the total number of sites with an ICS of $\nu$ in the SVT, where $N_0$ is the number of sites with only non-ionizing energy deposits. $N_{\nu,j}$ was obtained by

$$N_{\nu,j} = \frac{1}{n_t\, n_r} \sum_{t=1}^{n_t} \sum_{r=1}^{n_r} \sum_{k=\nu}^{\infty} \frac{N_{\nu,j}(k,t,r)}{k}, \tag{9}$$

Where $n_t$ is the number of simulated tracks, $n_r$ is the number of repetitions of the AVC algorithm, and $N_{\nu,j}(k,t,r)$ is the number of sites (with ICS $\nu$ and an energy imparted falling into the $j^{\text{th}}$ energy bin) scored with the $t^{\text{th}}$ track in the $r^{\text{th}}$ repetition that contain $k$ ETPs.

The frequency distribution of the ratio of energy imparted and ICS (for ICS greater than zero) was also determined conditional on ICS and SVT diameter.

The ICS defined by Eq. (4) was determined according to Eq. (10).

$$p_1(\nu|d_s, r_p) = \frac{N_\nu}{\sum_{\mu=1}^{\infty} N_\mu} \tag{10}$$

This is a conditional ICS distribution and inherently independent of the size of the SVT (as long as the latter envelopes the track). In addition, unconditional ICS distribution in the SVT given by

$$P_\nu(d_s, r_p) = N_\nu \frac{V_s}{A_p\, L'} \left(1 - \delta_{\nu,0}\right) + \delta_{\nu,0} \left(1 - \left(\sum_{\mu=1}^{\infty} N_\mu\right) \frac{V_s}{A_p\, L'}\right) \tag{11}$$

were also determined. In Eq. (11), $V_s = {d_s}^3\pi/6$ denotes the volume of the site, $A_p = {r_p}^2\pi$ the cross-sectional area of the SVT and $L' = L - d_s$ its effective length.

Eventually, the ratio of the number of events per fluence $n/\Phi$ was determined according to the following rationale: Consider all ETPs of a track within an SVT. The union of the spheres of diameter $d_s$ around all ETPs contains all points in the SVT that have at least one ETP within a maximum distance of $d_s/2$. This union is the associated volume of the track and can be segmented into disjoint sub-volumes $V_{\nu,k}$ containing points that have $k$ EPTs at distances of maximum $d_s/2$ around them of which $\nu$ result from ionizations.

The average cross-section of the track's associated volume along the length $L$' of the SVT is given by the ratio of this volume to $L$'. The total volume of the spheres around the ETPs in the SVT is their number $N_\varepsilon$ times $V_s$. Owing to the overlaps between these spheres, which define regions with more than one ETP in their vicinity, the associated volume of the track is reduced by a factor accounting for these overlaps. This factor is the mean number $M_{1,\varepsilon}$ of ETPs per site, which can be calculated by

$$M_{1,\varepsilon} = \frac{\sum_{\nu=0}^{\infty} \sum_{k=\nu}^{\infty} k\, V_{\nu,k}}{\sum_{\nu=0}^{\infty} \sum_{k=\nu}^{\infty} V_{\nu,k}} = \frac{\sum_{\nu=0}^{\infty} \sum_{k=\nu}^{\infty} k\, N_{\nu,k}}{\sum_{\nu=0}^{\infty} \sum_{k=\nu}^{\infty} N_{\nu,k}} = \frac{N_\varepsilon}{\sum_{\nu=0}^{\infty} N_{\nu,k}} \tag{12}$$

The number of events per fluence were determined by Eq. (13) for values of $k = 0$ (corresponding to the event definition in microdosimetry), $k = 1$ (corresponding to a nanodosimetric event as defined in Eq. (4)), and $k = 2$ (corresponding to 'true' ionization clusters as opposed to single ionizations).

$$\left(\frac{n}{\Phi}\right)_k = \sum_{\nu=k}^{\infty} N_\nu \times \frac{V_s}{L'} \tag{13}$$

## 3 Results and discussion

### *3.1 Dependence of the event frequency on beam size*

Fig. 4 shows the number of sites per length found on average around a proton track depending on the radius of the SVT used for scoring. The data refers to sites of 3 nm diameter and the length of the SVT cylinder was 997 nm, that is the full length of the proton trajectory segment minus the site diameter. The four panels of the plot relate to different energies of the protons (1 MeV, 10 MeV, 30 MeV, and 99 MeV). The data marked by squares are obtained when all sites with any (non-zero) value of energy imparted are scored, which corresponds to the definition of a microdosimetric event. The circles show the results when only sites containing at least one ionization are counted, corresponding to a nanodosimetric event as per Eq. (4). The asymptotic values of the number of sites per track, which is proportional to the expected number of events in a site irradiated by a proton beam, show large differences by a factor of 3, which appears to be independent on the proton energy.

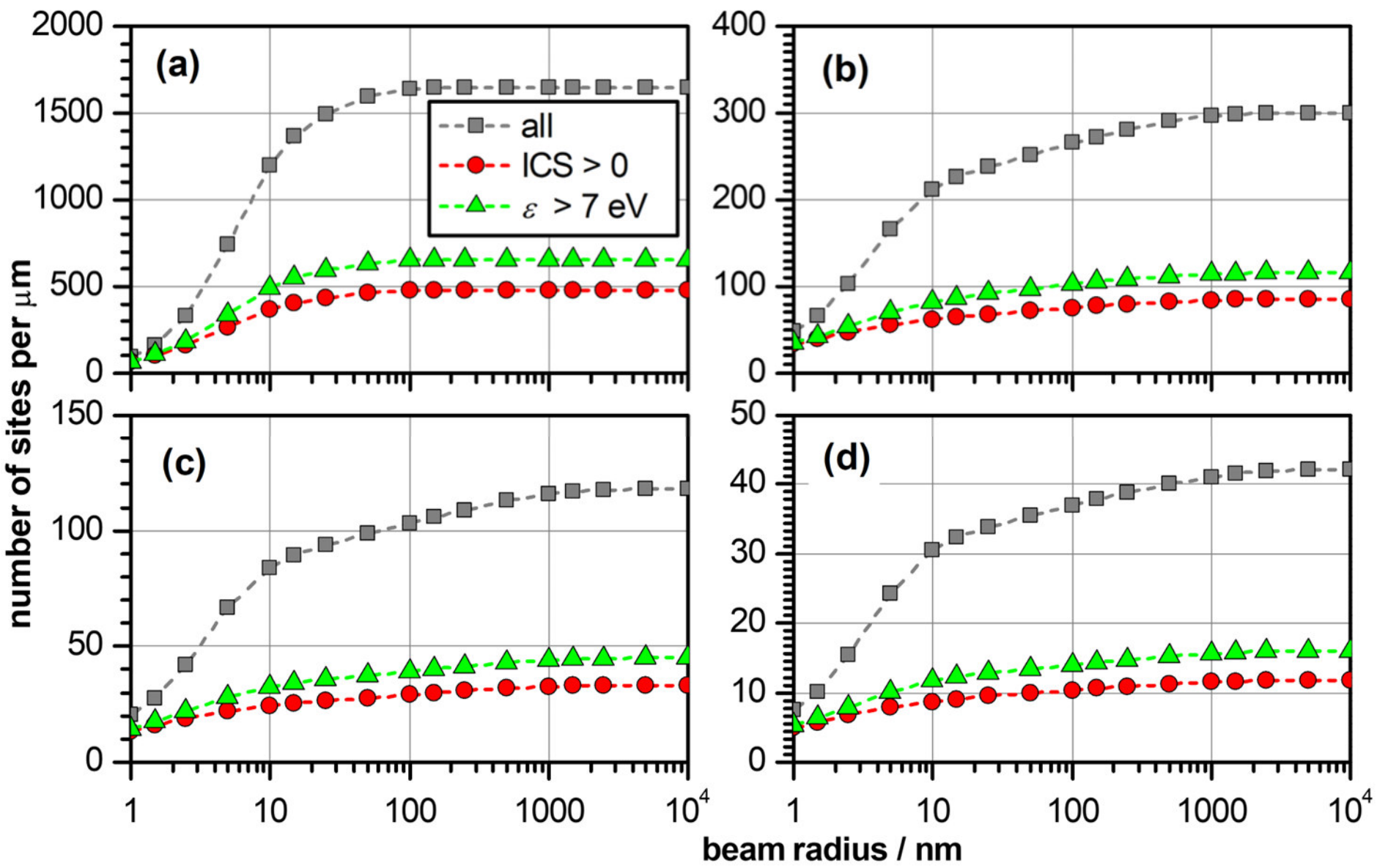

**Fig. 4**: Dependence of the scored number of 3 nm-diameter sites on the radius of the SVT for sites with any non-zero value of energy imparted (squares), with energy imparted exceeding 7 eV (triangles), and with at least one ionization (circles) for different proton energies of (a) 1 MeV, (b) 10 MeV, (c) 30 MeV, (d) 99 MeV. The data shown is the number of sites per path length of the proton track with site centers in a cylinder of the beam radius around the proton trajectory. The simulations were performed using Geant4-DNA option 2.

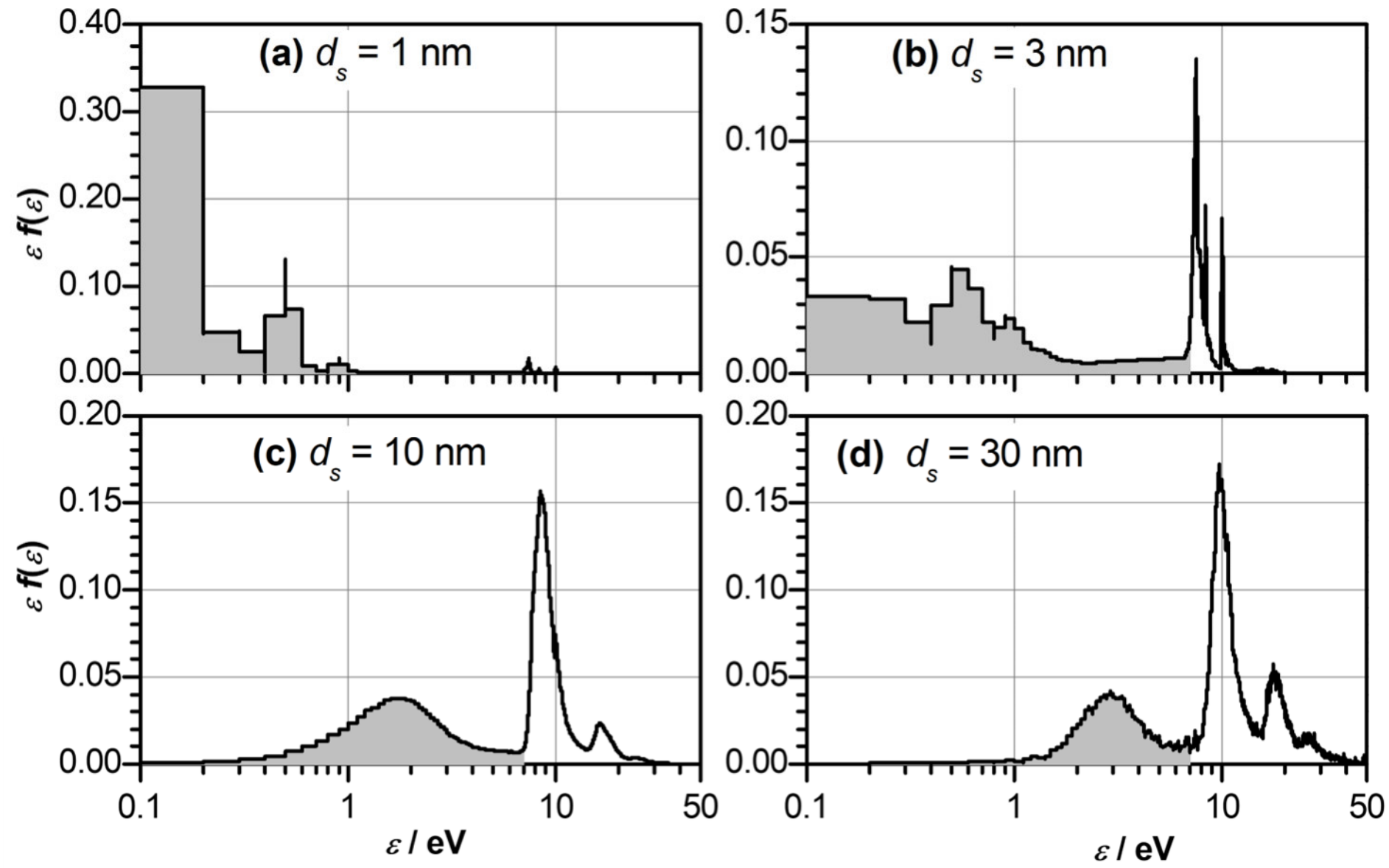

**Fig. 5**: Frequency distribution of energy imparted in sites in which no ionization occurred for a 1 MeV proton beam and site diameters of (a) 1 nm, (b) 3 nm, (c) 10 nm, (d) 30 nm. The data relate to simulations using Geant4-DNA option 2. The gray shaded area indicated the energy range below 7 eV.

The reason for this large discrepancy can be found in Fig. 5 which shows the frequency distribution of energy in sites without ionizations. Most of such sites receive very low values of energy imparted. This is because in these simulations (using Geant4 option2), electrons were followed until their energy fell below 1 eV and energy loss by vibrational excitations was included in the simulation which was the dominant energy loss process for sub-ionization electrons. At the same time, the corresponding cross-sections have low values resulting in a mean free path between inelastic collisions that allow the energy deposits to occur at positions well separated from the last ionization occurring (as compared with the site size).

The presentation of the data in Fig. 5 is in the conventional presentation style of microdosimetry, such that the area under the plotted curve per logarithmic interval is proportional to the contribution of the corresponding energy range to the total integral over the distribution. Therefore, Fig. 5 (a) shows that for the smallest considered site diameter of 1 nm, most sites without ionization have an energy imparted below 1 eV. With a site diameter of 3 nm, to which the results shown in Fig. 4 relate, Fig. 5 (b) shows that about 80 % of the sites have values of energy imparted below 7 eV, that is, below the pronounced peaks corresponding to electronic excitations. (There are only three peaks visible because autoionization was assumed for the other two possible excitations with energies above the ionization threshold.)

The proportion of sites with energy imparted in this energy range (gray shaded areas in Fig. 5) appears to be independent on the proton energy (Supplementary Fig. S1). On the other hand, it is worth noting that this proportion is much smaller with Geant4-DNA options 4 and 6 (Supplementary Fig. S2). This is a direct consequence of the fact that in these versions of Geant4-DNA the only non-ionizing energy loss process modelled is electronic excitations. Values of energy imparted smaller than the first excitation energy are therefore only from secondary electrons that are not transported in the simulation owing to their energy being below the set threshold. Therefore, the results obtained with Geant4-DNA option 2 shown in in Fig. 4, Fig. 5, and Supplementary Fig. S1 may be considered more realistic as regards the abundance of sites with energy deposits without ionization. For larger site sizes, energy imparted below 1 eV occurs less frequently, and the relative proportion of sites with energy imparted below 7 eV decreases (Fig. 5 (c) and (d)).

While break-up of molecular bonds is not expected to occur for vibrational excitations of the electronic ground state, electronic transitions to unoccupied molecular orbitals in the Frack-Condon regime may end up in a highly excited vibronic state that may lead to the break of a bond (Wang et al., 2020). Therefore, it is reasonable to exclude sites with energy imparted below the first electronic excitation energy. This has been done by counting only sites with energy imparted exceeding 7 eV, shown as triangles in Fig. 4. The corresponding number of sites per track only slightly exceeds the number of sites with ionizations.

This distinction between events fulfilling the formal definition (energy is imparted in the site) and events involving at least one ionization or at least one electronic excitation is only relevant for nanometric site sizes. With increasing site size, the proportion of sites with energy transfer points but without ionizations is rapidly decreasing, although even for the largest site size considered in this study of 100 nm, this proportion ranges between a few percent at low proton energies and up to about 30 % at the highest investigated proton energy of 99 MeV (Supplementary Fig. 3). The fraction of sites without ionization in which an energy imparted of 7 eV or higher is obtained is on the order of 50 % for options 4 and 6 of Geant4-DNA, whereas for option 2 a distinct dependence on site size is observed (Supplementary Fig. 4). This different behavior is again the consequence of which non-ionizing energy transfer processes are considered in the modelling.

The key message to be taken from Fig. 4, Fig. 5, and Supplementary Figs. S1 to S4 is the following: With nanometric targets, the event concept of conventional microdosimetry implies that a large proportion of the sites undergoing an event receives neither an ionization nor an electronic excitation. That is, a large proportion of the events have no biological relevance. Beyond this, it is worth noting that a microdosimetric measurement generally relies on the production of free charges by the events to be detected. Therefore, while the formal definition of an event in microdosimetry requires sites with energy imparted to be included even if no ionization occurs, it makes sense to use the concept of a nanodosimetric event also for microdosimetry with targets of sizes in the nanometer range. Therefore, in the following only sites with ionization will be considered.

### *3.2 Dependence of event frequency on site size*

Fig. 6 shows the variation of the proportion of sites with an ionization with the radius of the SVT for different site sizes and the same four proton energies used in Fig. 4. For 1 MeV protons (Fig. 6(a)), the proportion of sites that is found within 1 nm from the proton trajectory is about 5 % for the smaller site diameters up to 10 nm, and about 10 % for 30 nm and 100 nm diameter. Saturation of the site score is achieved within a few 100 nm from the proton trajectory. For

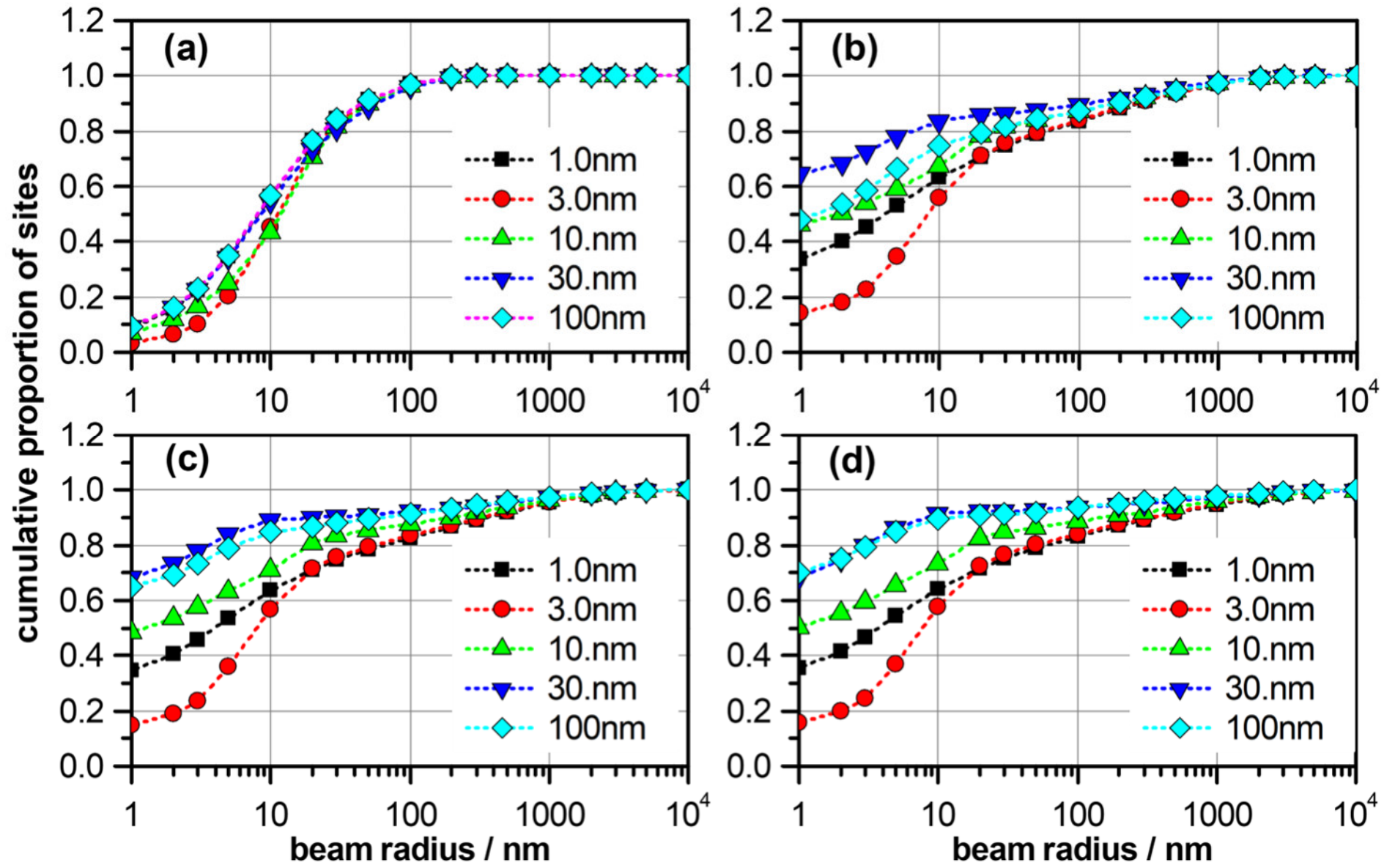

**Fig. 6**: Relative variation of the number of sites with ionizations as a function of the SVT radius for different site diameters and proton energies of (a) 1 MeV, (b) 10 MeV, (c) 30 MeV, (d) 99 MeV. The data shown were derived from the proton track simulations using Geant4-DNA option 2.

10 MeV protons, shown in Fig. 6(b), the SVT diameter for which saturation occurs is close to 1 µm and even larger values are found for the other two energies of 30 MeV and 99 MeV (Fig. 6(c) and (d)). Furthermore, at the higher energies shown in Fig. 6(b) to (d), larger values are found at the smallest SVT radius, and there is a larger spread between the data relating to different site diameters.

That the larger sites show higher values of the proportion of sites scored within smaller radial distance from the primary particle trajectory is understandable, since it is the center of the site that must be in the cylindrical SVT of the respective radius. Thus, the total volume sampled for ETPs by all possible sites within an SVT has a radius which is larger than the radius of the SVT by the site radius.

However, the variation with site diameter is not monotonous. For 10 MeV and 30 MeV (Fig. 6(b) and (c)), the curves for 30 nm site diameter have higher values at small beam radii than the curves pertaining to a 100 nm site diameter. Sites with 3 nm diameter have smaller values at small beam radii than sites with 1 nm diameter in all panels of Fig. 6. It should be noted that the error bars indicating the sampling statistics are not visible since they are smaller than the symbols. Therefore, these unexpected observations are not an effect of statistics.

The strongest variation with beam size is observed for the lowest proton energy of 1 MeV shown in Fig. 6(a). This appears surprising at first glance because this radiation quality is the most densely ionizing. However, the shorter mean free path for ionization by the proton also implies that the penumbra of ETPs due to electron interactions also is more densely populated since the distance between different electrons spurs decreases. In consequence the volume defined by all points of an SVT that are surrounded by ETPs can be expected to be only slightly dependent on the site diameter.

For higher proton energies, the curves shown in Fig. 6 indicate a rather complex dependence on primary particle energy of the spatial pattern of ETPs around a PPT. This spatial distribution is an interplay between the increase of the mean distance between successive interactions of the primary particle and the fact that most secondary electrons are produced by the electromagnetic pulse the ionized molecule experiences when a swift charged particle passes by. Only a small proportion of secondary electrons originate in binary collisions. Therefore, the energy distribution of emitted electrons is almost invariant with the primary particle energy, and so is the pattern of ETPs in an electron spur. The overall pattern of ETPs is the superposition of the ETPs of the primary particle and those from electrons, where different spurs interfere less with each other when the ETPs of the primary particle have larger separation at higher primary particle energies.

When sites of increasing size are used to sample the spatial pattern of ETPs, an initial decrease is expected. This is because sites with very small diameters score mostly single ETPs whereas with growing site size more and more sites will score several ETPs. This appears to be what can be seen for the 1 nm and 3 nm diameter sites and may be relevant for models connecting nanodosimetry with radiobiological effectiveness. A 3 nm sphere has the same volume as a cylinder representing a DNA segment of 10 base pairs which is often considered for such models (Grosswendt, 2006; Alexander et al., 2015a; Ramos-Méndez et al., 2018). On the other hand, empirically a 1 nm site diameter seems to give the best agreement between nanodosimetric quantities and radiobiological inactivation cross-sections (Conte et al., 2024, 2023). It is remarkable that there always seems to be convergence in Fig. 6 between the curves for these two sites sizes at about 10 nm to 20 nm from the PPT. It is worth noting that this convergence is also seen in the simulations using the other options of Geant4 included in this study (see Supplementary Figs. S5 and S6). However, for options 4 and 6, the differences between the two smallest site sizes are less prominent than for option 2.

### *3.3 Event horizon*

The results shown in Fig. 6 inform the question of how large a beam size must be chosen in a simulation to ensure that the ICS and imparted energy distributions determined are not impaired by an improper choice of this parameter. The corresponding data obtained from the simulation with Geant4-DNA option2 are presented in Fig. 7.

Fig. 7 shows the cylinder radius needed to score at least 98 % of the sites when applying the microscopic event definition (squares) or when requiring at least one ionization in the site. The two definitions essentially give the same beam radius. This radius increases with proton energy as expected. For all four site sizes shown in Fig. 7, the relevant beam radius is in the order of 20 nm at 1 MeV proton energy and increases up to about 300 nm at 10 MeV. Starting from about 40 MeV proton energy, different behavior is seen for the three smaller site sizes (Fig. 7(a) to (c)) and the 30 nm site diameter data in Fig. 7(d). For the smaller site sizes the radius within which 98 % of the sites are found almost doubles compared to what one would expect by extrapolation of the data at lower energy. This is a direct consequence of the fact that in Fig. 6 the curves for site diameters up to 10 nm have visible lower values than the curve for 30 nm.

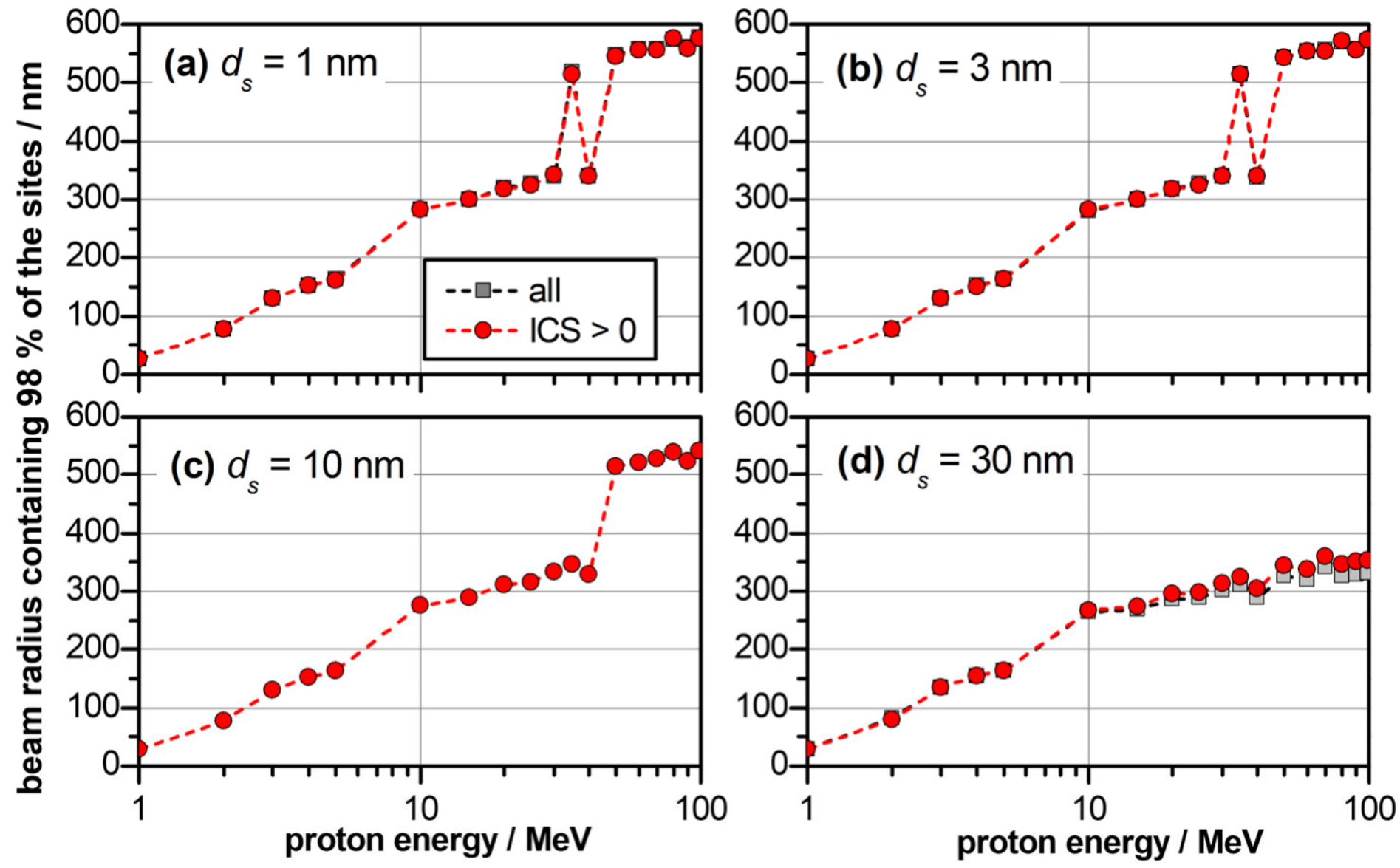

**Fig. 7**: Dependence on proton energy of the radii of cylinders around the proton track containing 98 % of the sites receiving any value of energy imparted (squares) or at least one ionization for different site diameter (a) 1 nm, (b) 3 nm, (c) 10 nm, (d) 30 nm calculated with Geant4-DNA option 2. In (a) to (c) the symbols of the two datasets coincide at all energies. The bi-stable behavior seen in (a) and (b) is due to the coarse grid of beam radii employed (of only 4 points per decade).

The radial range within which ionizations or energy deposits are found is expected to increase with proton energy owing to the increasing energy of electrons produced in binary encounter collisions. However, the step-like increase as well as the bi-stable behavior of the data seen in Fig. 7(a) and (b) at 35 MeV and 40 MeV may be artefacts related to the fact that only four beam radii per decade were considered in the analysis between which linear interpolation was applied. The absence of this apparent step with 30 nm site size and larger (not shown) can be attributed to the fact that the energy transfer points are sparser at larger distances from the PPT while for small SVT radii, a larger proportion of the SVT is filled with sites when the site size increases.

The beam radius containing 98 % of the sites appears to strongly depend on the cross-section data sets used in the simulations. The corresponding results for options 4 and 6 are shown in Supplementary Figs. S7 and S8. It can be seen in Supplementary Fig. S7 that for option 4 the 98 % radius is slightly smaller than for option 2 at proton energies below 10 MeV and become almost independent of proton energy for energies exceeding 10 MeV. For option 6 (Supplementary Fig. S8), on the other hand, the radii are comparable to those from option 2. This agrees with the fact that the electron cross-section for higher energy electrons in option 6 are identical to those in option 2.

The estimated radius of a cylinder within which 98 % of the energy transfer points or ionizations lie is an indication of the range of impact parameters that contribute to the energy imparted or ICS in a site. Similar to what was reported earlier (Braunroth et al., 2020), this relevant beam radius can amount to several hundreds of nanometers.

### *3.4 Fluence corresponding to a single event on average.*

As discussed in Section 2.1.5, the key quantity for determining multi-event distributions is the fluence $\phi_1$ producing one event on average. The corresponding results are shown in Fig. 8. Fig. 8(a) displays the variation with proton energy for different site sizes; Fig. 8 (b) shows the variation with site size for different proton energies. The three different symbols represent the data obtained with Geant4-DNA options 2 (full symbols), 4 (symbol with a cross) and 6 (half-filled symbols). The results for each option of Geant4-DNA are shown individually in Supplementary Figs. S9, S10, and S11, where data for all investigated site sizes are shown. For better comparison of the curves, the product of the fluence $\phi_1$ and the cross-sectional area $A_s$ of the site is shown on the $y$-axis. This quantity is directly interpretable as it is the number of primary particles traversing the cross section of the site.

It can be seen in Fig. 8 (a) that for the smaller site sizes shown (1 nm, 3 nm, and 10 nm) the number of tracks required to produce one event on average increases with increasing proton energy. For 1 nm and 3 nm, the corresponding datapoints appear to lie on an almost straight line in the log-log-plot. That is, the data follows a power law. For the two larger site sizes, the number of tracks is almost independent of proton energy for proton energies up to 10 MeV, and this number increases at higher proton energies. The slopes (or asymptotic slope for the 10 nm, 30 nm and 100 nm cases) appear to be similar, which suggests the same exponent of the power law (of about 0.8). This reflects the power-law decrease of the cross-section for proton-impact ionization. The almost constant number of tracks per single event for the larger sites at smaller proton energies is because there is almost always an interaction of protons of these energies in sites of these dimensions. It is worth noting, however, that this constant number is about 1.5 which seems to correspond to the ratio 3/2 of the diameter of the sphere to its mean chord length.

Fig. 8 (b) shows that the variation with site size seems to follow a power law with negative exponent for the two higher proton energies shown (30 MeV and 99 MeV). Corresponding to the analog observation of similar slopes at high energies in Fig. 8 (a), here a similar slope is seen for small site sizes (in the log-log plot). For the 1 MeV, 3 MeV and 10 MeV data, the number of tracks per site cross-section corresponding to one event converges to the value of 1.5 for larger site diameters. It is interesting that there appears to be a clear minimum around or slightly below unity at site sizes between 5 nm and 10 nm for the 1 MeV protons. It is speculated that these observations will presumably be qualitatively similar for tracks of heavier ions albeit with different values for the number of tracks per site cross-section leading to an event on average. This will be investigated in future studies.

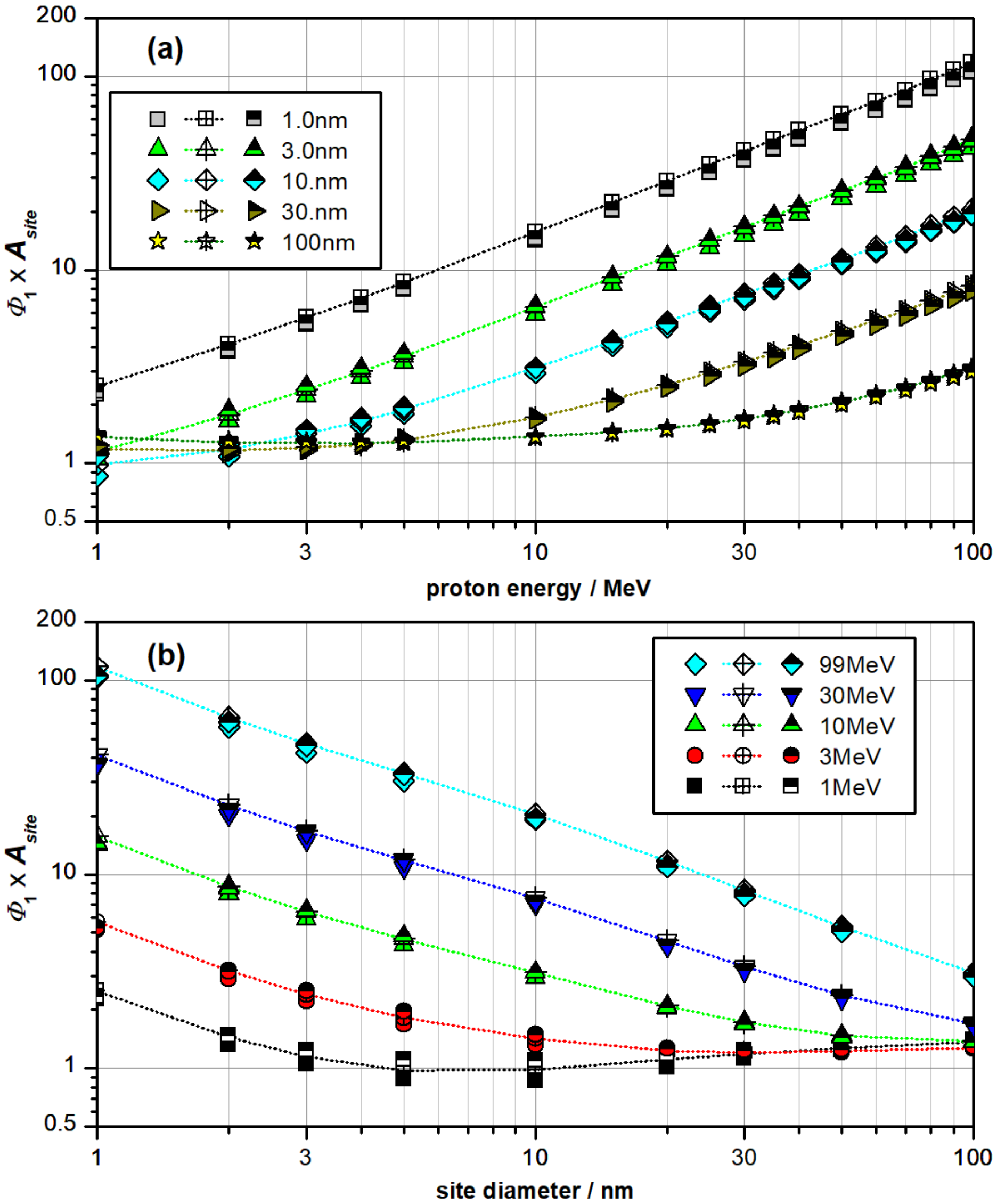

**Fig. 8**: Variation of the product of the fluence that produces an event on average and the cross-section of the site (a) with proton energy for different site diameters (see legend) and (b) with site diameter for several proton energies (see legend). The different symbol styles refer to simulations using Geant4-DNA options 2 (full symbols), 4 (open symbols with a crosshair) and 6 (symbols with upper half filled in black).

It is noted that the different versions of Geant4-DNA show similar trends in Fig. 8, while the absolute values show relative differences between the Geant4-DNA options in the order of 10 %. This may be interpreted as an indication of the uncertainty of multi-event microdosimetric or nanodosimetric descriptors of track structure and radiation quality. This uncertainty contribution is by a factor of 4 larger than the uncertainties acceptable for dosimetry in conventional radiotherapy (Nath et al., 1994). From previous investigations into the variation of results for different codes (Villagrasa et al., 2019), it may be expected that this uncertainty estimate becomes larger when results from simulations using other track structure codes would be included. This conjecture is confirmed by results of a recent code intercomparison exercise which also shows that the differences in nanodosimetric simulations results are to a major part due to the different models for electron interaction cross-sections (Villagrasa et al., 2025). On the other hand, the uncertainties related to biological weighting factors as used, for instance, in heavy ion therapy, may be of an even larger magnitude.

## 4 Conclusions

In conventional microdosimetry and nanodosimetry different concepts of what constitutes an event are used. While conventional nanodosimetry considers the coincidence between the passage of a primary ionizing particle and the occurrence of a number of ionizations (including none) in a site, conventional microdosimetry considers only PPTs that lead to energy deposits in the site. Conditional ICS distributions in nanodosimetry are somewhat analogous to the microdosimetric approach. However, with nanometric sites, the two event concepts remain disjoint even if conditional distributions in nanodosimetry are considered. This is because of the occurrence of sites in which only non-ionizing inelastic interactions occur.

While this was already indicated by results reported previously (Lillhök et al., 2022), the data presented here suggest that up to two thirds of the 3 nm diameter sites with energy deposits around proton tracks do not contain ionizations. On the other hand, about 80 % of such sites containing energy deposits exceeding the energy threshold of the first electronic excitation of the water molecule also contain an ionization. While it must be emphasized that these proportions depend strongly on the option of Geant4-DNA used for the simulations, the general trend is the same for all.

In the theoretical part, it was shown that for the conceptional gaps between microdosimetry and nanodosimetry for the same nanometric target to be overcome three key elements seem to be required. First, defining an event as relating to a beam volume within which interactions of the primary particle can result in energy deposits or ionizations in the site. Second, the use of conditional nanodosimetric ionization cluster size distributions. Third, the restriction of microdosimetric events by requiring the event to contain at least one ionization and, thus, be measurable by present-state microdosimetric counters. The implied unified event concept is a prerequisite for relating nanodosimetric quantities to fluence.

The data presented here further suggest that the range of impact parameters for which tracks can contribute to energy imparted and ionizations in a site may be hundreds of nm larger than the site size. This is not only the case when sites scoring ionizations are considered (Braunroth et al., 2020, 2021) but also for sites per the definition of microdosimetry. Conversely, the primary particle fluence producing one event on average was shown to have a non-trivial dependence on site size and radiation quality. This implies *inter alia* that the construction of muti-event distributions from measured or simulated single-event distributions cannot simply be achieved based on an event frequency estimated by multiplying the fluence of primary particles by the cross-section of the site. Therefore, the findings presented here are relevant for the potential future application of miniaturized nanodosimetric detectors in clinical practice with hadron therapy. This also applies to already existing miniaturized microdosimeters simulating nanometric sites as well as potential future solid-state microdosimeters featuring nanometric dimensions.

It should be noted that from the side of methodology, the concepts and data produced in this study can be employed to address further questions that were outside the scope of this paper. One is the variation of the means and other parameters of microdosimetric and nanodosimetric distributions within the BVS. Considerable variation would imply either using weighting approaches or reviewing the concepts of microdosimetry and nanodosimetry towards considering potential statistical distributions of those integral parameters. Clustering of sites should be another topic worth closer analysis.

Another interesting challenge could be to study which results would be obtained for the fluence producing one event on average when other radiation modalities are used, for instance, photons or electrons. A major challenge with such sparsely ionizing particles will be that the dimensions of the BVS will increase to macroscopic dimensions if high energy particles are considered as used in conventional radiotherapy. For low energy electrons, the challenge will be more that the assumption of straight particle trajectories fails, and alternative approaches must be conceived, e.g. using spherical BVS.

The considerations presented here appear highly relevant for the theoretical framework proposed recently for the use of nanodosimetric quantities in particle therapy treatment planning (Faddegon et al., 2023). Further development in future work is warranted to explore how the proposed event concept performs with other ion types and very densely ionizing particles.

## Acknowledgements

This work originated in the task on micro- and nanodosimetry within working group 6 “Computational Dosimetry” of the European Radiation Dosimetry Group (EURADOS eV). Anna Selva and Valeria Conte are acknowledged for stimulating discussions by which they contributed significantly to the foundations on which this work is built. LT acknowledges support from the “Metrology for Artificial Intelligence in Medicine (M4AIM)” program, funded by the German Federal Ministry of Economic Affairs and Climate Action in the frame of the QI-Digital Initiative.

## References

Alexander, F., Villagrasa, C., Rabus, H., Wilkens, J., 2015a. Local weighting of nanometric track structure properties in macroscopic voxel geometries for particle beam treatment planning. Phys. Med. Biol. 60, 9145–9156. https://doi.org/10.1088/0031-9155/60/23/9145

Alexander, F., Villagrasa, C., Rabus, H., Wilkens, J., 2015b. Energy dependent track structure parametrisations for protons and carbon ions based on nanometric simulations. Eur. Phys. J. D 69, 216-1-216–7. https://doi.org/10.1140/epjd/e2015-60206-5

Allisy, A., Jennings, W.A., Kellerer, A.M., Müller, J.W., 1993. ICRU Report 51: Quantities and Units in Radiation Protection Dosimetry. Journal of the ICRU os-26, iii–19.

Amols, H.I., Wuu, C.S., Zaider, M., 1990. On Possible Limitations of Experimental Nanodosimetry. Radiation Protection Dosimetry 31, 125–128. https://doi.org/10.1093/oxfordjournals.rpd.a080651

Bantsar, A., Colautti, P., Conte, V., Hilgers, G., Pietrzak, M., Pszona, S., Rabus, H., Selva, A., 2018. State of The Art of Instrumentation in Experimental Nanodosimetry. Radiat. Prot. Dosim. 180, 177–181. https://doi.org/10.1093/rpd/ncx263

Bernal, M.A., Bordage, M.C., Brown, J.M.C., Davídková, M., Delage, E., El Bitar, Z., Enger, S.A., Francis, Z., Guatelli, S., Ivanchenko, V.N., Karamitros, M., Kyriakou, I., Maigne, L., Meylan, S., Murakami, K., Okada, S., Payno, H., Perrot, Y., Petrovic, I., Pham, Q.T., Ristic-Fira, A., Sasaki, T., Štěpán, V., Tran, H.N., Villagrasa, C., Incerti, S., 2015. Track structure modeling in liquid water: A review of the Geant4-DNA very low energy extension of the Geant4 Monte Carlo simulation toolkit. Phys. Medica 31, 861–874. https://doi.org/10.1016/j.ejmp.2015.10.087

Besserer, J., Schneider, U., 2015. A track-event theory of cell survival. Z. Med. Phys. 25, 168–175. https://doi.org/10.1016/j.zemedi.2014.10.001

Bianchi, A., Colautti, P., Conte, V., Selva, A., Agosteo, S., Bortot, D., Mazzucconi, D., Pola, A., Reniers, B., Parisi, A., Struelens, L., Vanhavere, F., Tran, L., Rosenfeld, A.B., Cirrone, G.A.P., Petringa, G., 2020. Microdosimetry at the 62 MeV Proton Beam of CATANA: preliminary comparison of three detectors. Journal of Physics: Conference Series 1662, 012006. https://doi.org/10.1088/1742-6596/1662/1/012006

Bianchi, A., Mazzucconi, D., Selva, A., Colautti, P., Parisi, A., Vanhavere, F., Reniers, B., Conte, V., 2021. Lineal energy calibration of a mini-TEPC via the proton-edge technique. Radiat. Meas. 106526. https://doi.org/10.1016/j.radmeas.2021.106526

Bianchi, A., Selva, A., Eliasson, L., Lillhök, J., Rabus, H., Conte, V., 2024. Microdosimetry at the nanometre level with different techniques. Radiat. Phys. Chem. 221, 111778. https://doi.org/10.1016/j.radphyschem.2024.111778

Booz, J., Braby, L., Coyne, J., Kliauga, P., Lindborg, L., Menzel, H.-G., Parmentier, N., 1983. ICRU Report 36: Microdosimetry. Journal of the ICRU os-19, iii–119.

Bortot, D., Mazzucconi, D., Bonfanti, M., Agosteo, S., Pola, A., Pasquato, S., Fazzi, A., Colautti, P., Conte, V., 2017. A novel TEPC for microdosimetry at nanometric level: response against different neutron fields. Radiation Protection Dosimetry 180, 172–176. https://doi.org/10.1093/rpd/ncx198

Bortot, D., Mazzucconi, D., Fazzi, A., Agosteo, S., Pola, A., Colautti, P., Selva, A., Conte, V., 2022. From micro to nanodosimetry with an avalanche-confinement TEPC: Characterization with He-4 and Li-7 ions. Radiat. Phys. Chem. 198, 110225. https://doi.org/10.1016/j.radphyschem.2022.110225

Braby, L.A., Conte, V., Dingfelder, M., Goodhead, D.T., Pinsky, L.S., Rosenfeld, A.B., Sato, T., Waker, A.J., Guatelli, S., Magrin, G., Menzel, H.-G., Brandan, M.-E., Olko, P., 2023. ICRU Report 98, Stochastic Nature of Radiation Interactions: Microdosimetry. J. ICRU 23, 1–168. https://doi.org/10.1177/14736691231211380

Braunroth, T., Nettelbeck, H., Ngcezu, S.A., Rabus, H., 2021. Corrigendum to 'Three-dimensional nanodosimetric characterisation of proton track structure' [Radiation Physics and Chemistry 176 (2020) 109066]. Radiat. Phys. Chem. 186, 109535. https://doi.org/10.1016/j.radphyschem.2021.109535

Braunroth, T., Nettelbeck, H., Ngcezu, S.A., Rabus, H., 2020. Three-dimensional nanodosimetric characterisation of proton track structure. Radiat. Phys. Chem. 176, 109066. https://doi.org/10.1016/j.radphyschem.2020.109066

Bug, M.U., Baek, W.Y., Rabus, H., 2012. Simulation of ionisation clusters formed in nanometric volumes of the deoxyribose-substitute tetrahydrofuran. International Journal of Radiation Biology 88, 137–142. https://doi.org/10.3109/09553002.2011.610864

Casiraghi, M., Schulte, R., 2015. Nanodosimetry-based plan optimization for particle therapy. Computational and Mathematical Methods in Medicine 2015, 1–13. https://doi.org/10.1155/2015/908971

Conte, V., Bianchi, A., Rabus, H., Selva, A., 2024. Nanodosimetry applied to aerobic and hypoxic cells. Radiation Physics and Chemistry 111835. https://doi.org/10.1016/j.radphyschem.2024.111835

Conte, V., Bianchi, A., Selva, A., 2023. Track Structure of Light Ions: The Link to Radiobiology. IJMS 24, 5826. https://doi.org/10.3390/ijms24065826

Conte, V., Colautti, P., Grosswendt, B., Moro, D., Nardo, L.D., 2012. Track structure of light ions: experiments and simulations. New Journal of Physics 14, 093010. https://doi.org/10.1088/1367-2630/14/9/093010

Conte, V., Selva, A., Colautti, P., Hilgers, G., Rabus, H., 2017. Track structure characterization and its link to radiobiology. Radiat. Meas. 106, 506–511. https://doi.org/10.1016/j.Radmeas.2017.06.010

Conte, V., Selva, A., Colautti, P., Hilgers, G., Rabus, H., Bantsar, A., Pietrzak, M., Pszona, S., 2018. Nanodosimetry: towards a new concept of radiation quality. Radiat. Prot. Dosim. 180, 150–156. https://doi.org/10.1093/rpd/ncx175

De Nardo, L., Colautti, P., Conte, V., Baek, W.Y., Grosswendt, B., Tornielli, G., 2002. Ionization-cluster distributions of alpha-particles in nanometric volumes of propane: measurement and calculation. Radiat. Environ. Biophys. 41, 235–256. https://doi.org/10.1007/s00411-002-0171-6

Faddegon, B., Blakely, E.A., Burigo, L., Censor, Y., Dokic, I., Domínguez Kondo, N., Ortiz, R., Ramos Méndez, J., Rucinski, A., Schubert, K., Wahl, N., Schulte, R., 2023. Ionization detail parameters and cluster dose: a mathematical model for selection of nanodosimetric quantities for use in treatment planning in charged particle radiotherapy. Phys. Med. Biol. 68, 175013. https://doi.org/10.1088/1361-6560/acea16

Garty, G., Schulte, R., Shchemelinin, S., Grosswendt, B., Leloup, C., Assaf, G., Breskin, A., Chechik, R., Bashkirov, V., 2006. First attempts at prediction of DNA strand-break yields using nanodosimetric data. Radiation Protection Dosimetry 122, 451–454. https://doi.org/10.1093/rpd/ncl515

Garty, G., Schulte, R., Shchemelinin, S., Leloup, C., Assaf, G., Breskin, A., Chechik, R., Bashkirov, V., Milligan, J., Grosswendt, B., 2010. A nanodosimetric model of radiation-induced clustered DNA damage yields. Phys. Med. Biol. 55, 761–781. https://doi.org/10.1088/0031-9155/55/3/015

Goodhead, D.T., 2006. Energy deposition stochastics and track structure: what about the target? Radiat. Prot. Dosim. 122, 3–15. https://doi.org/10.1093/rpd/ncl498

Goodhead, D.T., 1994. Initial Events in the Cellular Effects of Ionizing Radiations: Clustered Damage in DNA. International Journal of Radiation Biology 65, 7–17. https://doi.org/10.1080/09553009414550021

Goodhead, D.T., 1989. The initial physical damage produced by ionizing radiations. Int J Radiat Biol 56, 623–634. https://doi.org/10.1080/09553008914551841

Grosswendt, B., 2006. Nanodosimetry, the metrological tool for connecting radiation physics with radiation biology. Radiat. Prot. Dosim. 122, 404–414. https://doi.org/10.1093/rpd/ncl469

Grosswendt, B., 2005. Nanodosimetry, from radiation physics to radiation biology. Radiat. Prot. Dosim. 115, 1–9. https://doi.org/10.1093/rpd/nci152

Grosswendt, B., 2002. Formation of ionization clusters in nanometric structures of propane-based tissue-equivalent gas or liquid water by electrons and α-particles. Radiat. Environ. Biophys. 41, 103–112. https://doi.org/10.1007/s00411-002-0155-6

Hawkins, R.B., 1998. A microdosimetric-kinetic theory of the dependence of the RBE for cell death on LET. AIP Conference Proceedings 25, 1157–1170.

Hilgers, G., Braunroth, T., Rabus, H., 2022. Correlated ionisations in two spatially separated nanometric volumes within the track structure of 241Am alpha particles: comparison with Monte Carlo simulations. Radiat. Phys. Chem. 201, 110488. https://doi.org/10.1016/j.radphyschem.2022.110488

Hilgers, G., Bug, M., Rabus, H., 2017. Measurement of track structure parameters of low and medium energy helium and carbon ions in nanometric volumes. Phys. Med. Biol. 62, 7569–7597.

Hilgers, G., Rabus, H., 2020. Correlated ionisations in two spatially separated nanometric volumes in the track structure of 241Am alpha particles: Measurements with the PTB ion counter. Radiat. Phys. Chem. 176, 109025. https://doi.org/10.1016/j.radphyschem.2020.109025

Incerti, S., Baldacchino, G., Bernal, M., Capra, R., Champion, C., Francis, Z., Guatelli, S., Guèye, P., Mantero, A., Mascialino, B., Moretto, P., Nieminen, P., Rosenfeld, A., Villagrasa, C., Zacharatou, C., 2010a. The Geant4-DNA project. Int. J. Model. Simul. Sci. Comput. 1, 157–178. https://doi.org/10.1142/S1793962310000122

Incerti, S., Ivanchenko, A., Karamitros, M., Mantero, A., Moretto, P., Tran, H.N., Mascialino, B., Champion, C., Ivanchenko, V.N., Bernal, M.A., Francis, Z., Villagrasa, C., Baldacchino, G., Guèye, P., Capra, R., Nieminen, P., Zacharatou, C., 2010b. Comparison of GEANT4 very low energy cross section models with experimental data in water. Med. Phys. 37, 4692–4708. https://doi.org/10.1118/1.3476457

Incerti, S., Kyriakou, I., Bernal, M.A., Bordage, M.C., Francis, Z., Guatelli, S., Ivanchenko, V., Karamitros, M., Lampe, N., Lee, S.B., Meylan, S., Min, C.H., Shin, W.G., Nieminen, P., Sakata, D., Tang, N., Villagrasa, C., Tran, H.N., Brown, J.M.C., 2018. Geant4-DNA example applications for track structure simulations in liquid water: A report from the Geant4-DNA Project. Med. Phys. 45, e722–e739. https://doi.org/10.1002/mp.13048

Kellerer, A.M., 1985. Fundamentals of Microdosimetry, in: The Dosimetry of Ionizing Radiation. Academic Press, pp. 77–162. https://doi.org/10.1016/B978-0-12-400401-6.50007-3

Lazarakis, P., Bug, M.U., Gargioni, E., Guatelli, S., Rabus, H., Rosenfeld, A.B., 2012. Comparison of nanodosimetric parameters of track structure calculated by the Monte Carlo codes Geant4-DNA and PTra. Phys. Med. Biol. 57, 1231. https://doi.org/10.1088/0031-9155/57/5/1231

Lea, D.E., 1955. Actions of Radiations on Living Cells, 2nd ed. University Press, Cambridge.

Lea, D.E., 1940. A radiation method for determining the number of genes in the X-chromosome of Drosophila. Journ. of Genetics 39, 181–188. https://doi.org/10.1007/BF02982834

Lillhök, J., Billnert-Maróti, R., Anastasiadis, A., 2022. MCNP 6.2 simulations of energy deposition in low-density volumes corresponding to unit-density volumes on the nanometre level. Radiat. Meas. 157, 106831. https://doi.org/10.1016/j.radmeas.2022.106831

Lindborg, L., Rabus, H., 2023. Technical note: Evaluation of a mean value of the number of ionizations in a single event distribution by the variance method in microdosimetry. Medical Physics 50, 5248–5251. https://doi.org/10.1002/mp.16428

Lindborg, L., Waker, A., 2017. Microdosimetry: Experimental Methods and Applications. CRC Press, Boca Raton.

Loncol, T., Cosgrove, V., Denis, J.M., Gueulette, J., Mazal, A., Menzel, H.G., Pihet, P., Sabattier, R., 1994. Radiobiological Effectiveness of Radiation Beams with Broad LET Spectra: Microdosimetric Analysis Using Biological Weighting Functions. Radiat. Prot. Dosim. 52, 347–352. https://doi.org/10.1093/rpd/52.1-4.347

Mazzucconi, D., Bortot, D., Pola, A., Agosteo, S., Selva, A., Colautti, P., Conte, V., 2020a. An Avalanche confinement TEPC as connecting bridge from micro to nanodosimetry. J. Phys.: Conf. Ser. 1662, 012023. https://doi.org/10.1088/1742-6596/1662/1/012023

Mazzucconi, D., Bortot, D., Rodriguez, P.M., Pola, A., Fazzi, A., Colautti, P., Conte, V., Selva, A., Agosteo, S., 2020b. A wall-less Tissue Equivalent Proportional Counter as connecting bridge from microdosimetry to nanodosimetry. Radiat. Phys. Chem. 171, 108729. https://doi.org/10.1016/j.radphyschem.2020.108729

Menzel, H.G., Pihet, P., Wambersie, A., 1990. Microdosimetric Specification of Radiation Quality in Neutron Radiation Therapy. International Journal of Radiation Biology 57, 865–883. https://doi.org/10.1080/09553009014550991

Mietelska, M., Pietrzak, M., Bancer, A., Ruciński, A., Szefliński, Z., Brzozowska, B., 2024. Ionization Detail Parameters for DNA Damage Evaluation in Charged Particle Radiotherapy: Simulation Study Based on Cell Survival Database. IJMS 25, 5094. https://doi.org/10.3390/ijms25105094

Nath, R., Biggs, P.J., Bova, F.J., Ling, C.C., Purdy, J.A., Van De Geijn, J., Weinhous, M.S., 1994. AAPM code of practice for radiotherapy accelerators: Report of AAPM Radiation Therapy Task Group No. 45. Medical Physics 21, 1093–1121. https://doi.org/10.1118/1.597398

Ngcezu, S., 2021. An investigation of nanodosimetric parameters for realistic geometries and a novel study of the track structure penumbra region (PhD). University of the Witwatersrand, Johannesburg, South Africa.

Ngcezu, S.A., Rabus, H., 2021. Investigation into the foundations of the track-event theory of cell survival and the radiation action model based on nanodosimetry. Radiat. Environ. Biophys. 60, 559–578. https://doi.org/10.1007/s00411-021-00936-4

Palmans, H., Rabus, H., Belchior, A., Bug, M., Galer, S., Giesen, U., Gonon, G., Gruel, G., Hilgers, G., Moro, D., Nettelbeck, H., Pinto, M., Pola, A., Pszona, S., Schettino, G., Sharpe, P., Teles, P., Villagrasa, C., Wilkens, J.J., 2015. Future development of biologically relevant dosimetry. Brit. J. Radiol. 88, 20140392. https://doi.org/10.1259/bjr.20140392

Parisi, A., Beltran, C.J., Furutani, K.M., 2022. The Mayo Clinic Florida microdosimetric kinetic model of clonogenic survival: formalism and first benchmark against in vitro and in silico data. Phys. Med. Biol. 67, 185013. https://doi.org/10.1088/1361-6560/ac7375

Pietrzak, M., 2019. On the two modes of nanodosimetric experiment. Radiat. Prot. Dosim. 183, 187–191. https://doi.org/10.1093/rpd/ncy233

Rabus, H., 2020. Nanodosimetry – on the “tracks” of biological radiation effectiveness. Z. Med. Phys. 30, 91–94. https://doi.org/10.1016/j.zemedi.2020.01.002

Rabus, H., Ngcezu, S.A., Braunroth, T., Nettelbeck, H., 2020. “Broadscale” nanodosimetry: Nanodosimetric track structure quantities increase at distal edge of spread-out proton Bragg peaks. Radiat. Phys. Chem. 166, 108515. https://doi.org/10.1016/j.radphyschem.2019.108515

Rabus, H., Palmans, H., Hilgers, G., Sharpe, P., Pinto, M., Villagrasa, C., Nettelbeck, H., Moro, D., Pola, A., Pszona, S., Teles, P., 2014. Biologically Weighted Quantities in Radiotherapy: an EMRP Joint Research Project. EPJ Web Conf. 77, 00021. https://doi.org/10.1051/epjconf/20147700021

Ramos-Méndez, J., Burigo, L.N., Schulte, R., Chuang, C., Faddegon, B., 2018. Fast calculation of nanodosimetric quantities in treatment planning of proton and ion therapy. Phys. Med. Biol. 63, 235015. https://doi.org/10.1088/1361-6560/aaeeee

Rossi, H.H., 1960. Spatial distribution of energy deposition by ionizing radiation. Radiation Research Supplement 2, 290–299. https://doi.org/10.2307/3583601

Rossi, H.H., Zaider, M., 1996. Microdosimetry and its Applications. Springer, Berlin, Heidelberg, New York. https://doi.org/10.1007/978-3-642-85184-1

Rucinski, A., Biernacka, A., Schulte, R., 2021. Applications of nanodosimetry in particle therapy planning and beyond. Phys. Med. Biol. 66, 24TR01. https://doi.org/10.1088/1361-6560/ac35f1

Schneider, U., Vasi, F., Schmidli, K., Besserer, J., 2020. A model of radiation action based on nanodosimetry and the application to ultra-soft X-rays. Radiat. Environ. Biophys. 59, 439–450. https://doi.org/10.1007/s00411-020-00842-1

Schneider, U., Vasi, F., Schmidli, K., Besserer, J., 2019. Track Event Theory: A cell survival and RBE model consistent with nanodosimetry. Radiat. Prot. Dosim. 183, 17–21. https://doi.org/10.1093/rpd/ncy236

Schulte, R., Bashkirov, V., Shchemelinin, S., Garty, G., Chechik, R., Breskin, A., 2001. Modeling of radiation action based on nanodosimetric event spectra. Phys Medica 17 Suppl 1, 177–180.

Schulte, R.W., Wroe, A.J., Bashkirov, V.A., Garty, G.Y., Breskin, A., Chechik, R., Shchemelinin, S., Gargioni, E., Grosswendt, B., Rosenfeld, A.B., 2008. Nanodosimetry-based quality factors for radiation protection in space. Z. Med. Phys. 18, 286–296. https://doi.org/10.1016/j.zemedi.2008.06.011

Selva, A., Bolst, D., Bianchi, A., Guatelli, S., Conte, V., 2023. Energy imparted and ionisation yield at the nanometre scale: results for extended beams. Radiat. Prot. Dosim. 199, 1984–1988. https://doi.org/10.1093/rpd/ncac253

Selva, A., Bolst, D., Guatelli, S., Conte, V., 2022a. Energy imparted and ionization yield in nanometre-sized volumes. Radiat. Phys. Chem. 192, 109910. https://doi.org/10.1016/j.radphyschem.2021.109910

Selva, A., Bolst, D., Guatelli, S., Conte, V., 2022b. Energy imparted and ionization yield in nanometre-sized volumes. Radiat. Phys. Chem. 192, 109910. https://doi.org/10.1016/j.radphyschem.2021.109910

Selva, A., Conte, V., Colautti, P., 2018. A Monte Carlo tool for multi-target nanodosimetry. Radiat. Prot. Dosim. 180, 182–186. https://doi.org/10.1093/rpd/ncy027

Selva, A., Nadal, V.D., Cherubini, R., Colautti, P., Conte, V., 2019. Towards the use of nanodosimetry to predict cell survival. Radiat. Prot. Dosim. 183, 192–196. https://doi.org/10.1093/rpd/ncy274

Thomas, L., Schwarze, M., Rabus, H., 2024. Radial dependence of ionization clustering around a gold nanoparticle irradiated by x-rays under charged particle equilibrium. Phys. Med. Biol. 69, 185014. https://doi.org/10.1088/1361-6560/ad6e4f

Villagrasa, C., Baiocco, G., Chaoui, Z.-E.-A., Dingfelder, M., Incerti, S., Kundrát, P., Kyriakou, I., Matsuya, Y., Kai, T., Paris, A., Perrot, Y., Pietrzak, M., Schuemann, J., Rabus, H., 2025. Evaluation of the uncertainty in calculating nanodosimetric quantities due to the use of different interaction cross sections in Monte Carlo track structure codes. arXiv:2502.11991 [physics.med-ph]. https://doi.org/10.48550/ARXIV.2502.11991

Villagrasa, C., Bordage, M.-C., Bueno, M., Bug, M., Chiriotti, S., Gargioni, E., Heide, B., Nettelbeck, H., Parisi, A., Rabus, H., 2019. Assessing the contribution of cross-sections to the uncertainty of Monte Carlo calculations in micro- and nanodosimetry. Radiat. Prot. Dosim. 183, 11–16. https://doi.org/10.1093/rpd/ncy240

Wang, E., Ren, X., Baek, W., Rabus, H., Pfeifer, T., Dorn, A., 2020. Water acting as a catalyst for electron-driven molecular break-up of tetrahydrofuran. Nat. Commun. 11, 2194. https://doi.org/10.1038/s41467-020-15958-7

## Supplement 1 - Appendix to "On a revised concept of an event that allows linking nanodosimetry and microdosimetry in nanometric sites with macroscopic dosimetry"

This Appendix presents theoretical considerations on how energy imparted or a certain number of ionizations in a site result from the superposition of charged particle tracks and a microscopic or nanoscopic site. For simplicity, only spherical sites and unidirectional irradiation are considered, and two approximations are used in the mathematical argument: first, the primary particles travel along straight lines in the beam volume of the site (BVS)[3]; second, their energy loss within the BVS is negligible compared to their average kinetic energy. Under these conditions, the volume density of possible energy transfer points with interactions of the primary particles is uniform in the BVS. This applies to single events as well as to multiple events and for any considered target volume within the BVS.

*A.1 Single events in microdosimetry*

The single event frequency distribution $f_1(\varepsilon)$ of energy imparted, $\varepsilon$, is the superposition of the contributions from all primary particle tracks that can lead to ETPs in the site. Consider a primary particle trajectory (PPT) along the unit vector $\hat{p}$ and let $\vec{r}_p$ be the vector perpendicular to $\hat{p}$ connecting the center of the site and the closest point to it on the PPT (cf. Fig. 3 in the main text). The length of vector $\vec{r}_p$ is the impact parameter.

If $F_1(\varepsilon|\vec{r},\vec{p})$ denotes the probability of such a PPT to produce an energy imparted up to $\varepsilon$ in the site, then the probability density $f_1^*$ is given by Eq. (A.1).

$$f_1^*(\varepsilon|\vec{r}_p,\hat{p}) = \frac{dF_1}{d\varepsilon}(\varepsilon|\vec{r}_p,\hat{p}) \tag{A.1}$$

It should be noted that there is the possibility of obtaining an energy imparted of $\varepsilon_0 = 0$ eV, and the corresponding probability is $F_0(\vec{r},\vec{p}) = F_1(\varepsilon_0|\vec{r},\vec{p})$. Therefore, if the irradiation was with this pencil beam, the single event frequency distribution $f_1(\varepsilon|\vec{r}_p,\hat{p})$ was given by Eq. (A.2).

$$f_1(\varepsilon|\vec{r}_p,\hat{p}) = \frac{f_1^*(\varepsilon|\vec{r}_p,\hat{p})}{1 - F_0(\vec{r},\hat{p})} \tag{A.2}$$

Considering now the BVS illustrated in Fig. 3(a), ´the probability density for obtaining an energy imparted $\varepsilon$ is $f_1^*(\varepsilon)$ given by Eq. (A.3).

$$f_1^*(\varepsilon) = \int\limits_{4\pi}\int\limits_{A_p} f_1^*(\varepsilon|\vec{r}_p,\hat{p})\frac{d^2\varphi(\vec{r}_p,\hat{p})}{dA d\Omega} d^2\vec{r}_p d^2\hat{p} \tag{A.3}$$

In Eq. (A.3), the inner integral extends over the cross-section of the BVS, $A_p$. $d^2\varphi/dAd\Omega$ is the probability density of the primary particle to have a direction of motion $\hat{p}$ and to pass at the point specified by $\vec{r}_p$. The specification of radiation quality (type and energy of the primary particle) is omitted for better legibility.

[3] See Fig. 3 in the paper and the related text. In brief, the BVS is the cylindrical envelope of the region around a site in which interactions of a primary particle producing a secondary electron have a non-zero probability of leading to energy deposits or ionizations in the site.

For a spherical site and uniform irradiation, the probability density $d^2\varphi/dAd\Omega$ is constant given by Eq. (A.4).

$$\frac{d^2\varphi\left(\vec{r}_p,\hat{p}\right)}{dAd\Omega} = \frac{1}{4\pi}\frac{1}{A_p} \tag{A.4}$$

Eq. (A.3) can then be simplified to the following expression for the single-event frequency distribution

$$f_1(\varepsilon) = \frac{1}{1-F_0}\frac{1}{A_p}\int_{A_p} f_1^*(\varepsilon|\vec{r})\, d^2\vec{r}\,, \tag{A.5}$$

where $F_0$ (Eq. (A.8)) is the probability of no energy being deposited in the target volume by the track, and $A_p$ is the cross-sectional area of the BVS. The factor $1/A_p$ represents a fluence of one particle crossing the area $A_p$, and $f_1^*(\varepsilon|\vec{r})$ (Eq. (A.7)) is the probability distribution of energy imparted for a primary particle trajectory passing the center of the target volume with a lateral offset $\vec{r}$.

It should be noted that Eq. (A.5) can also be written in terms of a weighted sum of the single-event frequency distributions from the different tracks (Eq. (A.6))

$$f_1(\varepsilon) = \frac{1}{1-F_0}\frac{1}{A_p}\int_{A_p} [1-F_0(\vec{r})] \times f_1(\varepsilon|\vec{r})\, d^2\vec{r}\ , \tag{A.6}$$

Without loss of generality, the direction of the PPT can be assumed to be along the *z* direction and the BVS to be confined by the planes $z = 0$ and $z = L$. The probability density $f_1^*(\varepsilon|\vec{r})$ of energy imparted for a primary particle trajectory passing the center of the target volume with a lateral offset given by $\vec{r}$ can be decomposed into components corresponding to a given number $k$ of interactions of the primary particle inside the BVS. The interaction of the primary particle at a point with coordinates $(z,\vec{r})$ may result in a contribution to the energy imparted in the site with a probability of $f_1^*(\varepsilon_i|z,\vec{r})d\varepsilon_i$ for this contribution to be between $\varepsilon_i$ and $\varepsilon_i+d\varepsilon_i$. To obtain the overall probability distribution $f_1^*(\varepsilon|\vec{r})$ for non-zero value of $\varepsilon$, the contributions from tracks with different number of interactions and from all possible distributions of the interaction points of the primary particles must be summed as stated by Eq. (A.7).

$$\begin{aligned} f_1^*(\varepsilon|\vec{r};Q,d_s) = \sum_{k=1}^{\infty}\left(\frac{\bar{k}_L}{L}\right)^k e^{-\bar{k}_L} \iiint_0^{\varepsilon} \int_0^L f_1^*(\varepsilon_1|z_1,\vec{r}) \int_{z_1}^L f_1^*(\varepsilon_2|z_2,\vec{r}) \cdots \\ \times \int_{z_{k-1}}^L f_1^*(\varepsilon_k|z_k,\vec{r}) dz_k \cdots dz_2\, dz_1 \times \delta\left(\varepsilon - \sum_i \varepsilon_i\right) d\varepsilon_1\, d\varepsilon_2 \cdots d\varepsilon_k \end{aligned} \tag{A.7}$$

Here $k$ and $\bar{k}_L$ are the actual and the average number of energy transfer points from interactions of the primary particle along the path of length $L$ occurring at the *z*-coordinates $z_1 \ldots z_k$. The quantities $\varepsilon_1 \ldots \varepsilon_k$ are the energies imparted in the target volume resulting from these interactions of the primary particle. These energies imparted should not be confused with energy deposits (energy transferred in a single interaction) and can also be zero. $f_1(\varepsilon_i|z,\vec{r})d\varepsilon_i$ is the probability of an interaction of the primary particle at the point with coordinates $(z,\vec{r})$ to result in an energy imparted in the target volume between $\varepsilon_i$ and $\varepsilon_i + d\varepsilon_i$, and δ is Dirac's delta function. Using δ avoids the clumsy notation of multiple convolutions since it assures that the energies imparted resulting from the interactions of the primary particle sum up to the considered value of energy imparted $\varepsilon$. The factor before the integrals in Eq. (A.7) is the probability of $k$ interactions according to a Poisson distribution with expectation $\bar{k}_L$.

$F_0$ in Eq. (A.5) is given by Eq. (A.8) where $\varepsilon_0$ = 0 eV and, contrary to Eq. (A.7), also the case $k$ = 0 is implicitly included.

$$F_0(Q, d_s, A_p) = \left(1 + \sum_{k=1}^{\infty} \left(\frac{\bar{k}_L}{L}\right)^k \frac{1}{A_p} \int_{A_p} F_1{}^{(k)}(\varepsilon_0|\vec{r}) d^2\vec{r}\right) e^{-\bar{k}_L} \tag{A.8}$$

with

$$F_1{}^{(k)}(\varepsilon_0|\vec{r}) = \int_0^L F_1(\varepsilon_0|z_1,\vec{r}) \int_{z_1}^L F_1(\varepsilon_0|z_2,\vec{r}) \cdots \int_{z_{k-1}}^L F_1(\varepsilon_0|z_k,\vec{r}) dz_k \cdots dz_2\, dz_1 \tag{A.9}$$

This probability $F_0(Q, d_s, A_p)$ in Eq. (A.8) is the sum of two contributions. One is the probability that the primary particle does not interact in the BVS. The second contribution is the sum of the probabilities of the primary particle undergoing $k$ interactions in the BVS of which none results in an energy deposit in the target volume. $F_1{}^{(k)}(\varepsilon_0|\vec{r})$ in Eq. (A.9) is the probability of all $k$ interactions of the primary particle in the BVS not leading to an energy deposit in the site, averaged over all possible distributions of the interactions along the stretch of the primary particle trajectory in the BVS.

### *A.2 Single events in nanodosimetry*

A notation analogous to that applied for microdosimetry in the previous section is also used here for the case of nanodosimetry, different from that conventionally used. See

Supplementary Table 1 for a comparison of symbols used here and in other work (De Nardo et al., 2002; Grosswendt, 2006). In the present notation, the ICS distribution is given by Eq. (A.10),

$$p_1(\nu|r_A; Q, d_s, A) = \frac{1}{A} \int_A p_1^{\#}(\nu|\vec{r})\, d^2\vec{r}\,, \tag{A.10}$$

where $r_A$ is the length of the vector $\vec{r}_A$ denoting the lateral offset between the centers of the site and the center of the area of the primary particle detector $A$. Given by Eq. (A.11), $p_1^{\#}(\nu|\vec{r})$ is the conditional probability of $\nu$ ionizations to occur in a site when a track passes at a point of closest approach which is offset from the site's center by the vector $\vec{r}$. The superscript "#" indicates that in this case the probability distribution also contains an entry for no ionizations occurring (ICS zero). (The parameters $Q$, $d_s$ and $A_p$ denoting the radiation quality, site diameter and cross-sectional area of the BVS, respectively, are omitted in Eq. (A.12) and on the right-hand side of Eq. (A.11) for better legibility.)

$$p_1^{\#}(\nu|\vec{r};Q,d_s) = \left(\delta_{\nu,0} + \sum_{k=1}^{\infty}\left(\frac{\bar{k}_L}{L}\right)^k p_1{}^{(k)}(\nu|\vec{r})\right)e^{-\bar{k}_L} \tag{A.11}$$

with

$$\begin{aligned} p_1{}^{(k)}(\nu|\vec{r}) = \sum_{\nu_1,\nu_2,\dots,\nu_k} \int_0^L p_1(\nu_1\,|z_1,\vec{r}) \int_{z_1}^L p_1(\nu_2\,|z_2,\vec{r}) \cdots \\ \times \int_{z_{k-1}}^L p_1(\nu_k\,|z_2,\vec{r}) dz_k \cdots dz_2\, dz_1\, \delta_{\nu,\sum_i \nu_i} \end{aligned} \tag{A.12}$$

The quantity $p_1{}^{(k)}(\nu|\vec{r})$ in Eq. (A.11) is the probability of obtaining $\nu$ ionizations in the site when there are $k$ interactions of the primary particle inside the BVS. The interaction of the primary particle at a point with coordinates $(z,\vec{r})$ contributes $\nu_i$ ionizations to the ICS in the site with a probability of $p_1(\nu_i\,|z,\vec{r})$. To obtain $p_1{}^{(k)}(\nu|\vec{r})$, contributions from all possible distributions of the interaction points of the primary particles must be added. That the sum of the contributed numbers of ionizations sums up to the required ICS $\nu$ is assured by $\delta_{i,j}$, Kronecker's delta symbol, which is unity when the two subscripts agree and else zero. As with Eq. (A.7), the use of $\delta_{i,j}$ replaces the conventionally employed formulation with multiple convolutions. To obtain the overall conditional frequency distribution $p_1^{\#}(\nu|\vec{r})$ , the contributions from tracks with different number of interactions must be summed as shown in Eq. (A.11).

It should be noted that, in simulation studies, nanodosimetric quantities are often determined according to Eq. (A.11), that is for an ideal pencil beam (Bug et al., 2012; Lazarakis et al., 2012; Conte et al., 2017, 2023). In contrast, conventional nanodosimetric experiments always relate to a primary particle detector of finite size such that an average over impact parameters according to Eq. (A.10) is determined. If scoring is performed in simulations by placing a number of target volumes around a primary particle track (Alexander et al., 2015b; Selva et al., 2018; Ramos-Méndez et al., 2018), the determined ICS distributions are also determined according to Eq. (A.10). This is so because the ionization cluster scored on a site is the result of the relative position of the track and the site. If the area $A$ used in Eq. (A.10) is not too large, the results obtained by Eqs. (A.11) and (A.10) are not too different. However, significantly different ICS distributions are obtained when a "broad beam" geometry is considered (Hilgers et al., 2017).

It should also be noted that Eq. (A.11) is more general than the formulas given in (De Nardo et al., 2002; Grosswendt, 2006) which appear to be based on the assumption that $p_1(\nu|z,\vec{r})$ is independent of $z$ within the SVT and zero outside. In which case one would obtain Eq. (A.13),where $\bar{p}_1(\nu_i\,|\vec{r})$ given by Eq. (A.14) is the probability of an interaction of the primary particle along the track segment of length $L$ to result in $\nu_i$ ionizations in the site. (Eq. (A.13) is the same as Eq. (4) in (De Nardo et al., 2002). The dependence of $\bar{p}_1$ on $\vec{r}$ has been omitted on the right-hand side of Eq. (A.13) to simplify notation.)

$$p_1^{\#}(\nu|\vec{r}) = \left( \delta_{\nu,0} + \sum_{k=1}^{\infty} \frac{\bar{k}_L{}^k}{k!} \sum_{\nu_1,\nu_2,\dots,\nu_k} \bar{p}_1(\nu_1)\bar{p}_1(\nu_2)\cdots\bar{p}_1(\nu_k)\delta_{\nu,\sum_i \nu_i} \right) e^{-\bar{k}_L} \tag{A.13}$$

with

$$\bar{p}_1(\nu_i|\vec{r}) = \int_0^L p_1(\nu_i|z,\vec{r})\, dz \tag{A.14}$$

However, even if the true $z$-dependence could be approximated by a constant along a stretch of the PPT and zero outside, it appears unlikely that the length of this stretch would be independent on $\nu$.

The structural similarity between Eqs. (A.7) and (A.11) suggests that replacing Eq. (A.10) with

$$p_1(\nu|r_A;Q,d_s,A) = \frac{1}{1 - P_{0,\mathrm{d}}(r_A;Q,d_s,A)} \frac{1}{A} \int_A p_1^{\#}(\nu|\vec{r})\, d^2\vec{r} \tag{A.15}$$

would assimilate nanodosimetric and microdosimetric distributions. The probability $P_{0,\mathrm{d}}$ of no ionization in the target volume when a primary particle trajectory passes through an area $A$ located at a lateral offset $\vec{r}_A$ relative to the target volume is given by Eq. (A.16).

$$P_{0,\mathrm{d}}(r_A;Q,d_s,A) = \left( 1 + \sum_{k=1}^{\infty} \left(\frac{\bar{k}_L}{L}\right)^k \frac{1}{A} \int_A p_1{}^{(k)}(0|\vec{r})\, d^2\vec{r} \right) e^{-\bar{k}_L} \tag{A.16}$$

with

$$p_1{}^{(k)}(0|\vec{r}) = \int_0^L p_1(0|z_1,\vec{r}) \int_{z_1}^L p_1(0|z_2,\vec{r}) \cdots \times \int_{z_{k-1}}^L p_1(0|z_k,\vec{r}) dz_k \cdots dz_2\, dz_1 \tag{A.17}$$

As with $F_0$, $P_{0,d}$ is the sum of the probability that there are no interactions of the primary particle in the BVS and the probabilities that the primary particle interacts $k$ times in the BVS but none of these interactions results in an ionization in the site. The quantity defined by Eq. (A.15) is referred to as conditional ICS distribution (Conte et al., 2012).

The ICS distributions in Eq. (A.15) are conditional both on the occurrence of at least one ionization in the site and on the passage of the primary particle through area $A$. However, it is also possible to consider the passage of the primary particle through the cross-sectional area $A_p$ of the BVS (Fig. 3(a)) and to define a single event ionization cluster size distribution $p_1(\nu)$ by

$$p_1(\nu|Q,d_s,A_p) = \frac{1}{1 - P_0} \frac{1}{A_p} \int_{A_p} p_1^{\#}(\nu|\vec{r})\, d^2\vec{r} \tag{A.18}$$

with the probability $P_0$ of no ionization in the target volume when a primary particle trajectory passes through the cross-section $A_p$ of the BVS given by Eq. (A.19).

$$P_0(Q,d_s,A_p) = \left( 1 + \sum_{k=1}^{\infty} \left(\frac{\bar{k}_L}{L}\right)^k \frac{1}{A_p} \int_{A_p} p_1{}^{(k)}(0|\vec{r}) d^2\vec{r} \right) e^{-\bar{k}_L} \tag{A.19}$$

Eq. (A.19) is the nanodosimetric analog to Eq. (A.8). The quantity $p_1{}^{(k)}(0|\vec{r})$ in Eq. (A.19) is given by

$$p_1{}^{(k)}(0|\vec{r}) = \int_0^L p_1(0|z_1,\vec{r}) \int_{z_1}^L p_1(0|z_2,\vec{r}) \cdots \times \int_{z_{k-1}}^L p_1(0|z_k,\vec{r}) dz_k \cdots dz_2\, dz_1 \qquad \text{(A.20)}$$

and represents the conditional probability that all $k$ interactions of the primary particle in the BVS do not lead to an ionization in the site, averaged over all possible distributions of the interactions along the stretch of the primary particle trajectory in the BVS.

In summary, Eq. (A.5) equates the single event frequency distribution of energy imparted to the multi-event frequency distribution obtained for a fluence of one primary particle per cross section $A_p$ of the BVS. Eq. (A.18) is the corresponding expression for the ionization cluster size distribution resulting from a fluence of this value, which can be interpreted as a redefined concept of a nanodosimetric event which is analogous to the microdosimetric event concept.

## Supplementary Table to “On a revised concept of an event that allows linking nanodosimetry and microdosimetry in nanometric sites with macroscopic dosimetry”

Supplementary Table 1: Comparison of the notation for nanodosimetric quantities used here to that used in earlier work by De Nardo et al. [1] and Grosswendt [2].

| De Nardo et al. [1] | Grosswendt [2] | This work | Meaning |
|---|---|---|---|
| $k$ | $Q$ | $Q$ | Radiation quality |
| $D$ | $D$ | $d_s$ | Site diameter |
| $d$ | $d$ | $r = \|\vec{r}\|$ | Impact parameter |
| $P_\nu(k; d)$ | $P_\nu(Q; d)$ | $p_1^{\#}(\nu\|\vec{r}; Q, d_s)$ | Ionization cluster size probability |
| $\bar{\kappa}(Q)$ | $\bar{\kappa}(Q)$ | $\bar{k}_L$ | Mean number of energy transfer points with ionizations of the primary particle in the beam volume relevant to the site |
| $f_\nu^{(1)}(d)$ | $f_\nu^{(1)}(d)$ | $\frac{1}{L}\int_0^L p_1(\nu\|z, \vec{r})dz$ | Probability of formation of an ionization cluster of size $\nu$, resulting from a single interaction of the primary particle along a stretch of length $L$ of its trajectory passing the site with an impact parameter $d$ (equal to the magnitude of the lateral offset $\vec{r}$). |

[1] De Nardo, L., Colautti, P., Conte, V., Baek, W.Y., Grosswendt, B., Tornielli, G., 2002. Ionization-cluster distributions of alpha-particles in nanometric volumes of propane: measurement and calculation. Radiation and Environmental Biophysics 41, 235–256. https://doi.org/10.1007/s00411-002-0171-6

[2] Grosswendt, B., 2006. Nanodosimetry, the metrological tool for connecting radiation physics with radiation biology. Radiat. Prot. Dosim. 122, 404–414. https://doi.org/10.1093/rpd/ncl469

# Supplementary Figures to “On a revised concept of an event that allows linking nanodosimetry and microdosimetry in nanometric sites with macroscopic dosimetry”

*Comparison of energy imparted in sites without ionization for different proton energies and different options of Geant4-DNA*

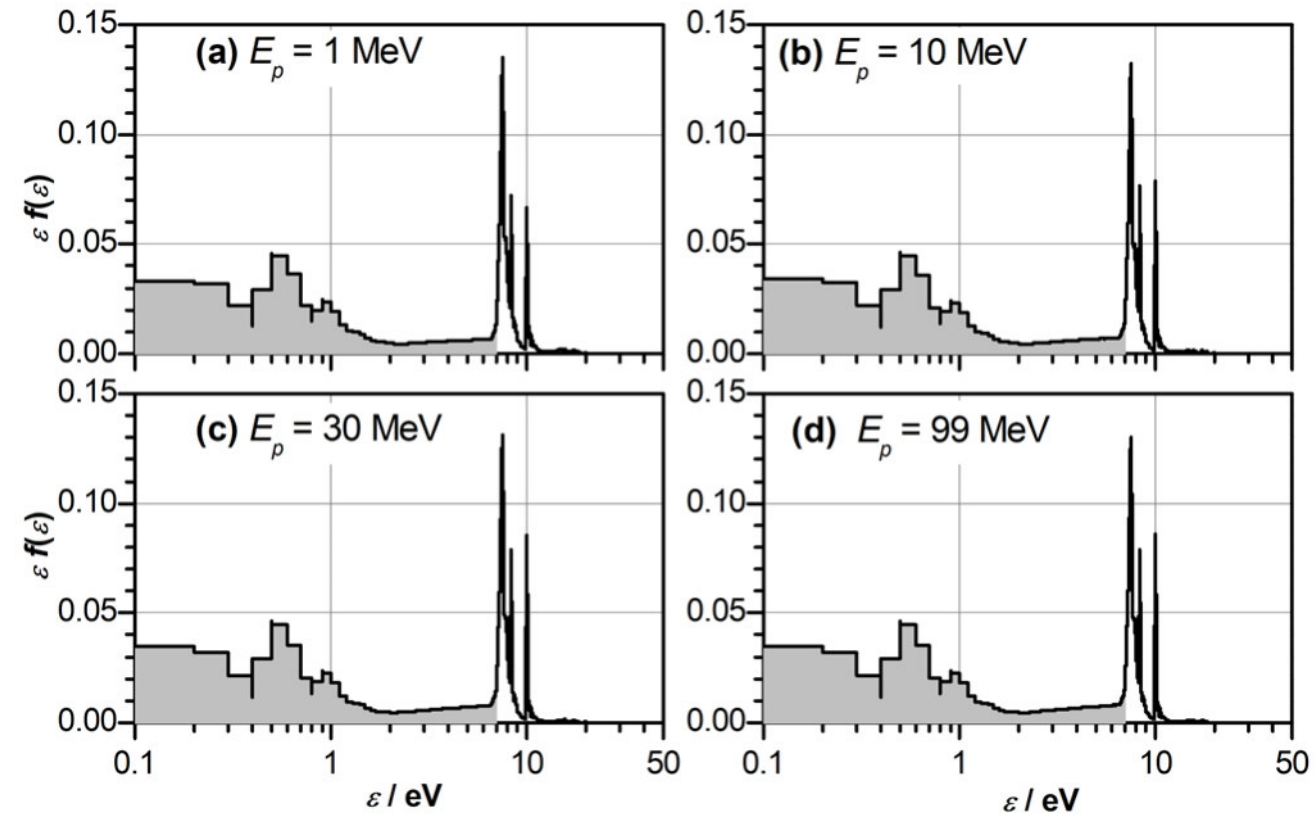

**Supplementary Fig. S1**: Comparison of the distributions of energy imparted in 3 nm diameter sites without ionization for different proton energies: (a) 1 MeV, (b) 10 MeV, (c) 30 MeV, and (d) 99 MeV. The simulations were performed using Geant4-DNA option 2.

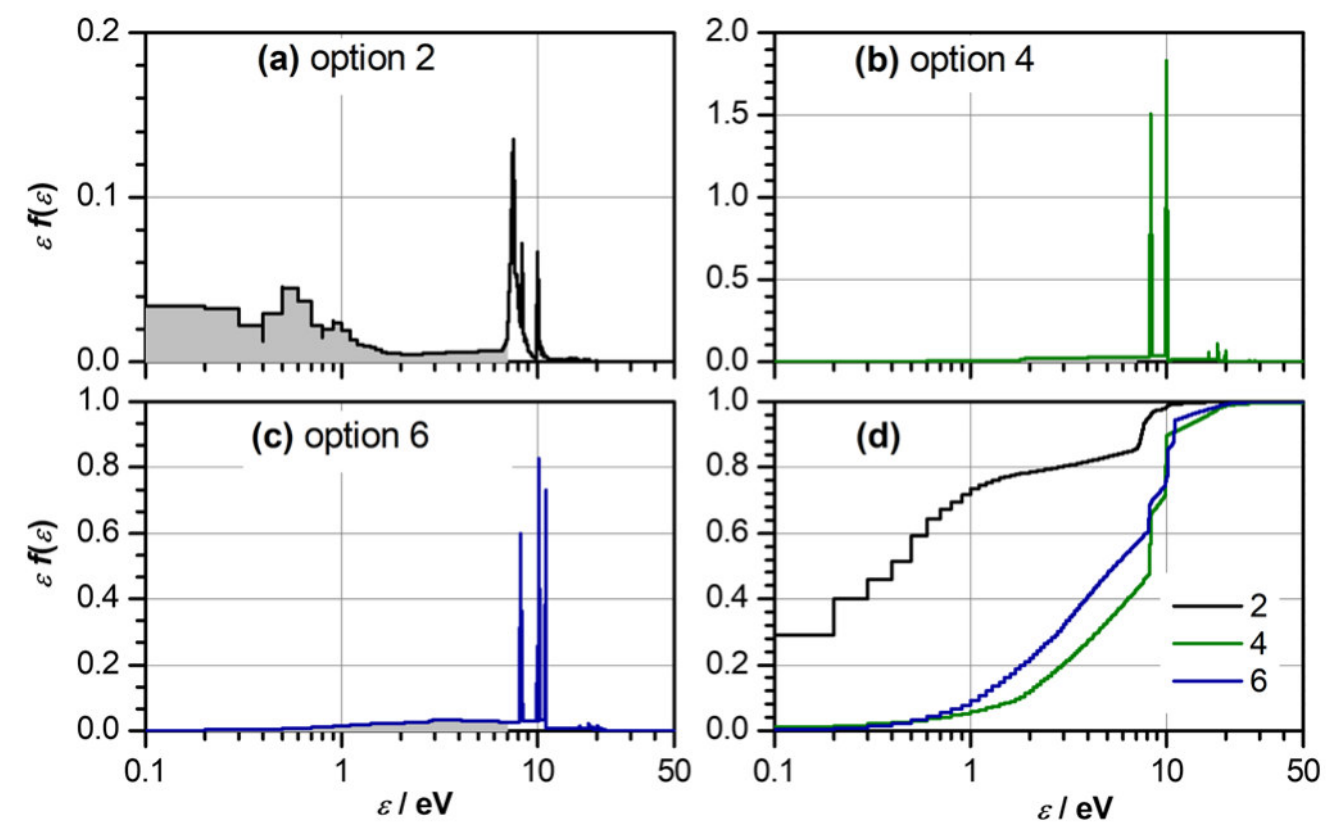

**Supplementary Fig. S2**: Comparison of the distributions of energy imparted in 3 nm diameter sites without ionization for 1 MeV proton beams obtained with Geant4-DNA option 2 (a), option 4 (b) and option 6 (c). (d) shows a comparison of the cumulative distributions. The peaks corresponding to Geant4-DNA options 4 and 6 are sharper because option 2 also considers interactions producing vibronic excitations which cause a broadening.

*Relevance of sites without ionizations*

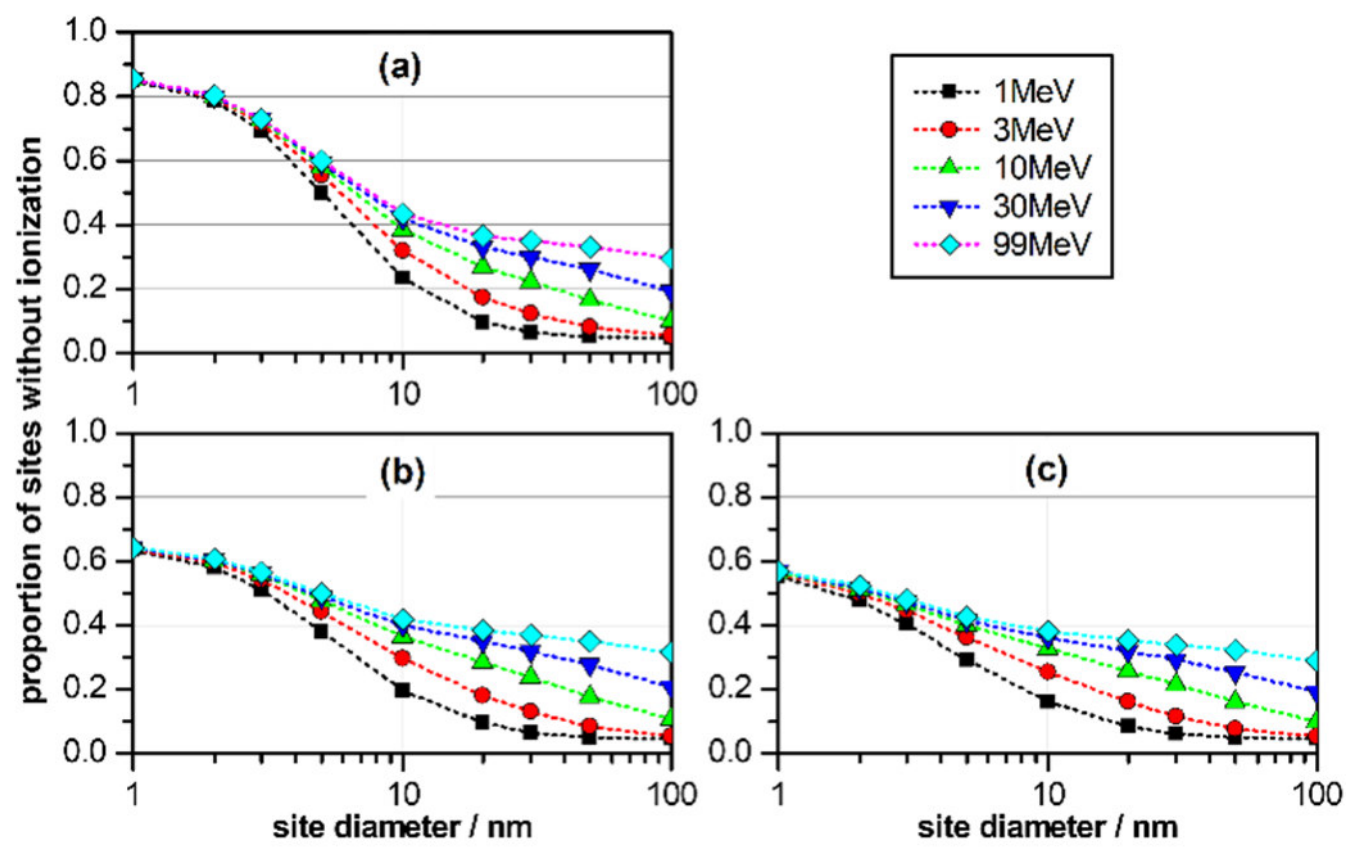

**Supplementary Fig. 3**: Proportion of sites around a proton trajectory in which no ionization occurs as a function of the site size for different proton energies (see legend) and simulations with (a) Geant4-DNA option 2, (b) Geant4-DNA option 4, (c) Geant4-DNA option 6.

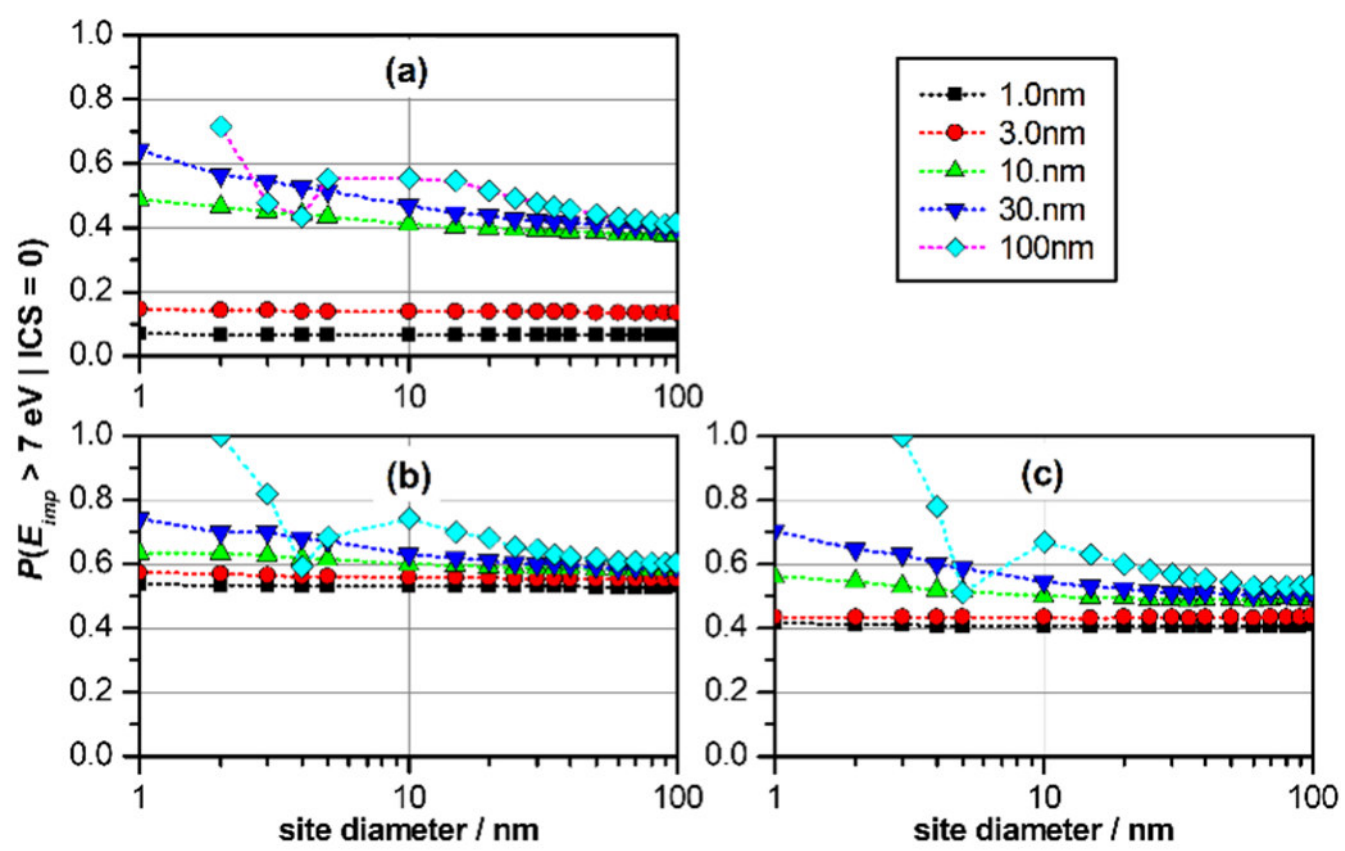

**Supplementary Fig. 4**: Proportion of sites around a proton trajectory without ionization that receive an energy imparted exceeding 7 eV as a function of the site size for different proton energies (see legend) and simulations with (a) Geant4-DNA option 2, (b) Geant4-DNA option 4, (c) Geant4-DNA option 6.

*Relative variation of the number of scored sites with beam radius for options 4 and 6*

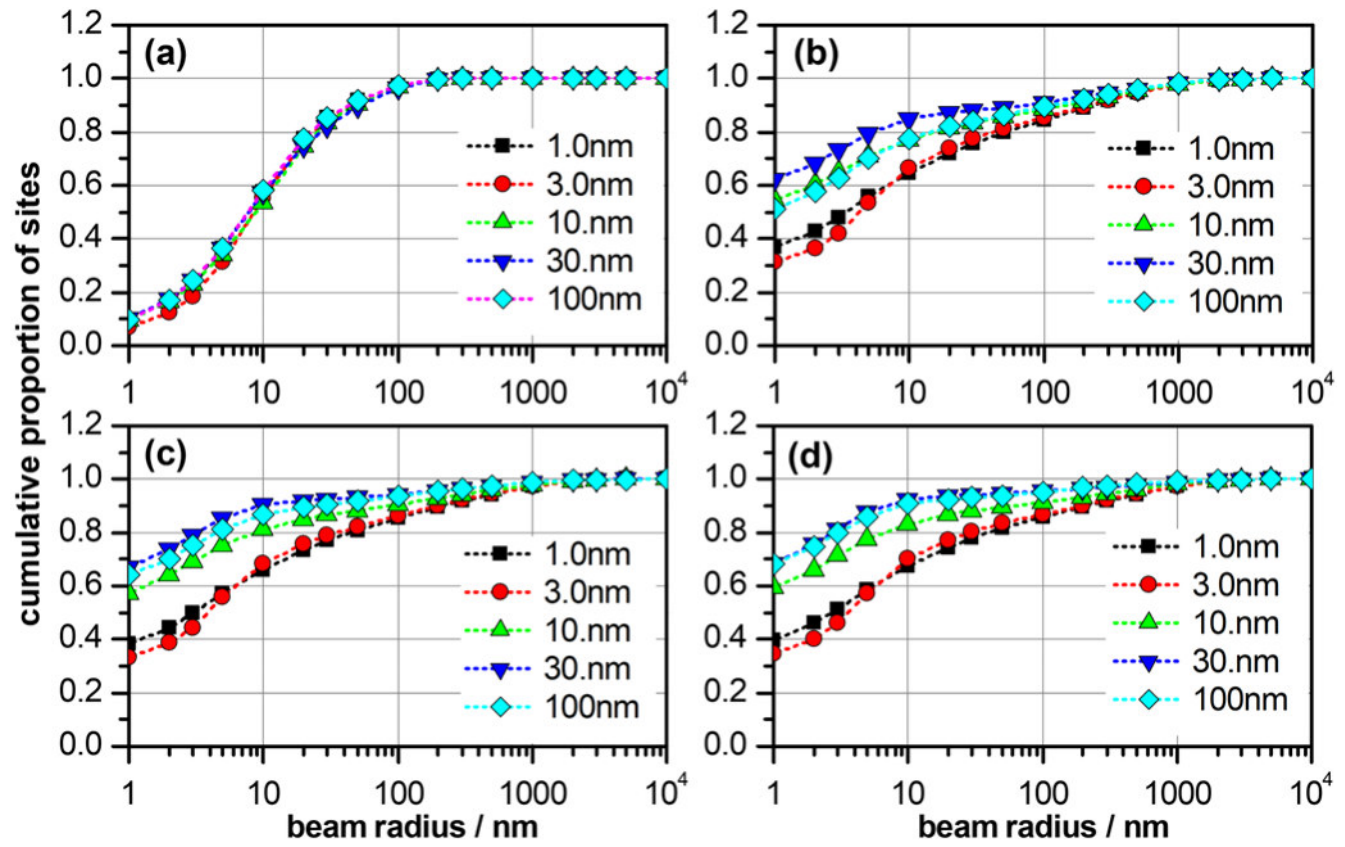

**Supplementary Fig. S5**: Same as Fig. 6 in the main text but for data obtained from the proton track simulations using Geant4-DNA option 4: Relative variation of the ratio of events per fluence for different site diameters and proton energies of (a) 1 MeV, (b) 10 MeV, (c) 30 MeV, (d) 99 MeV.

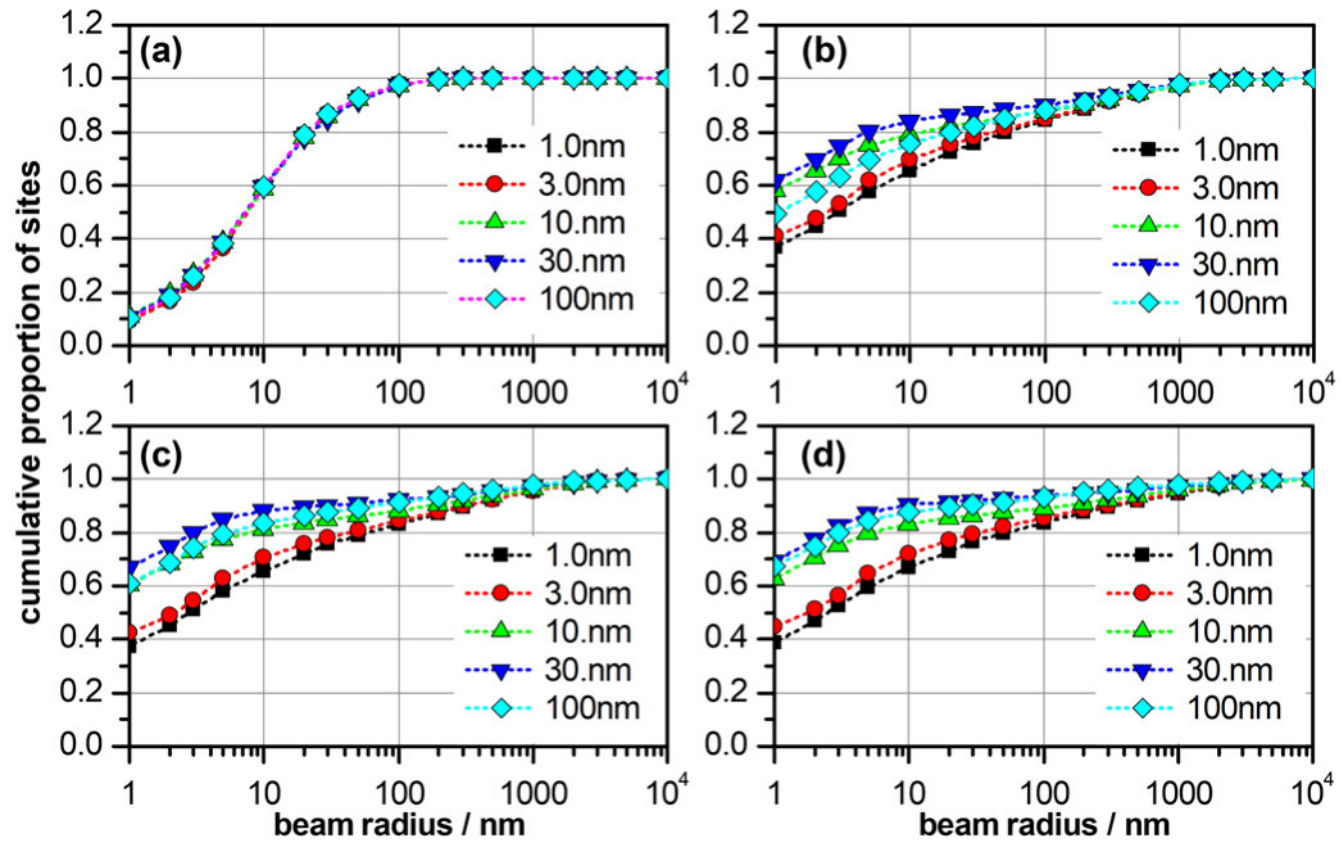

**Supplementary Fig. S6**: Same as Fig. 6 in the main text but for data obtained from the proton track simulations using Geant4-DNA option 6: Relative variation of the ratio of events per fluence for different site diameters and proton energies of (a) 1 MeV, (b) 10 MeV, (c) 30 MeV, (d) 99 MeV.

*Beam radius for Opt4 and Opt6*

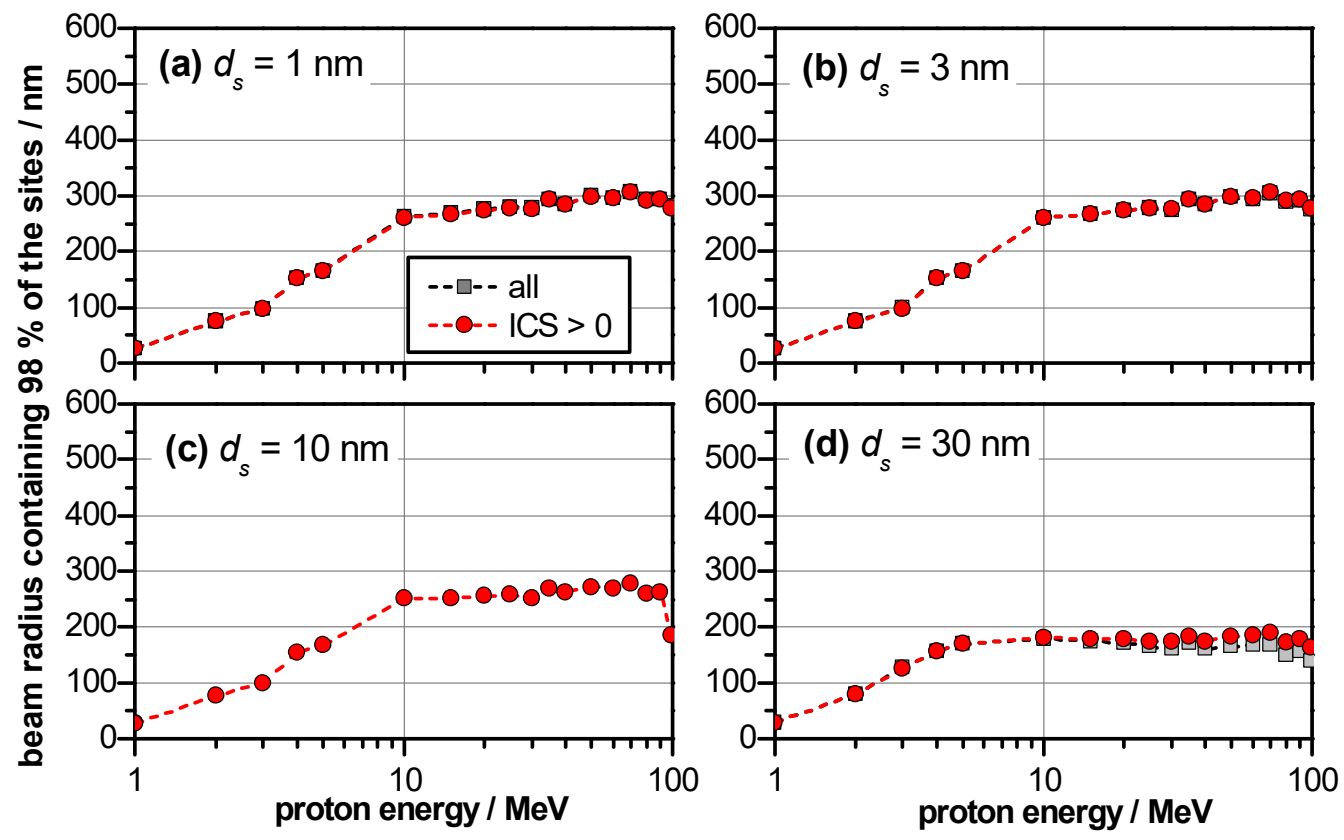

**Supplementary Fig. S7**: Same as **Fig. 7** in the main text but for data obtained from the proton track simulations using Geant4-DNA option 4: Dependence on proton energy of the radii of cylinders around the proton track containing 98 % of the sites receiving any value of energy imparted (squares) or at least one ionization for different site diameter (a) 1 nm, (b) 3 nm, (c) 10 nm, (d) 30 nm. In (a) to (c) the symbols of the two datasets coincide at all energies

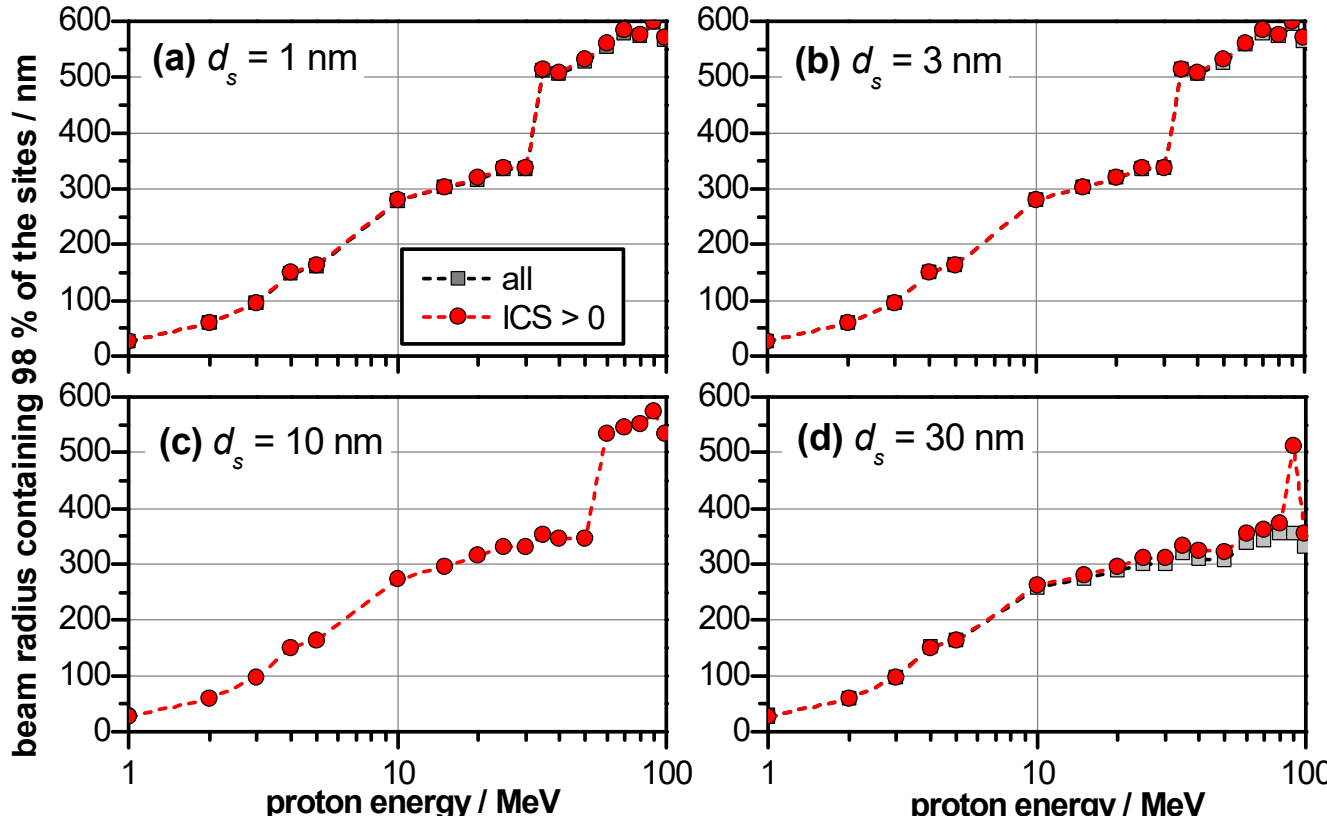

**Supplementary Fig. S8**: Same as **Fig. 7** in the main text but for data obtained from the proton track simulations using Geant4-DNA option 6: Dependence on proton energy of the radii of cylinders around the proton track containing 98 % of the sites receiving any value of energy imparted (squares) or at least one ionization for different site diameter (a) 1 nm, (b) 3 nm, (c) 10 nm, (d) 30 nm. In (a) to (c) the symbols of the two datasets coincide at all energies. The glitch seen in (d) is due to the coarse grid of beam radii employed (of only 4 points per decade).

*Comparison of the number of tracks producing one event on average obtained with the different options of Geant4-DNA*

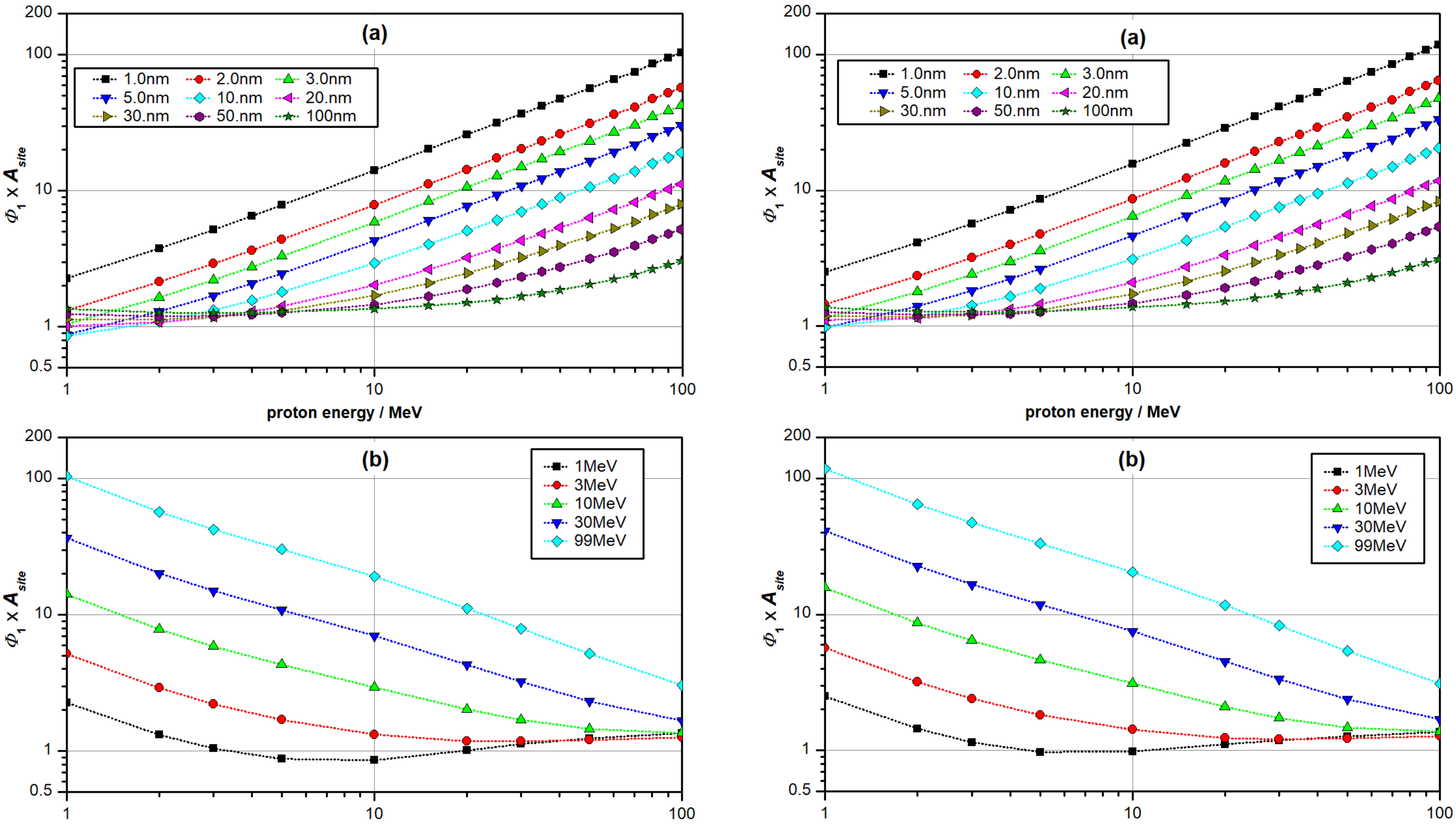

**Supplementary Fig. S9**: Dependence of the product between the fluence producing an event frequency of unity and the cross-section of the site (a) on proton energy for different site sizes (see legend) and (b) on site size for different proton energies (see legend). The data pertain to simulations performed with Geant4-DNA option 2.

**Supplementary Fig. S10**: Dependence of the product between the fluence producing an event frequency of unity and the cross-section of the site (a) on proton energy for different site sizes (see legend) and (b) on site size for different proton energies (see legend). The data pertain to simulations performed with Geant4-DNA option 4.

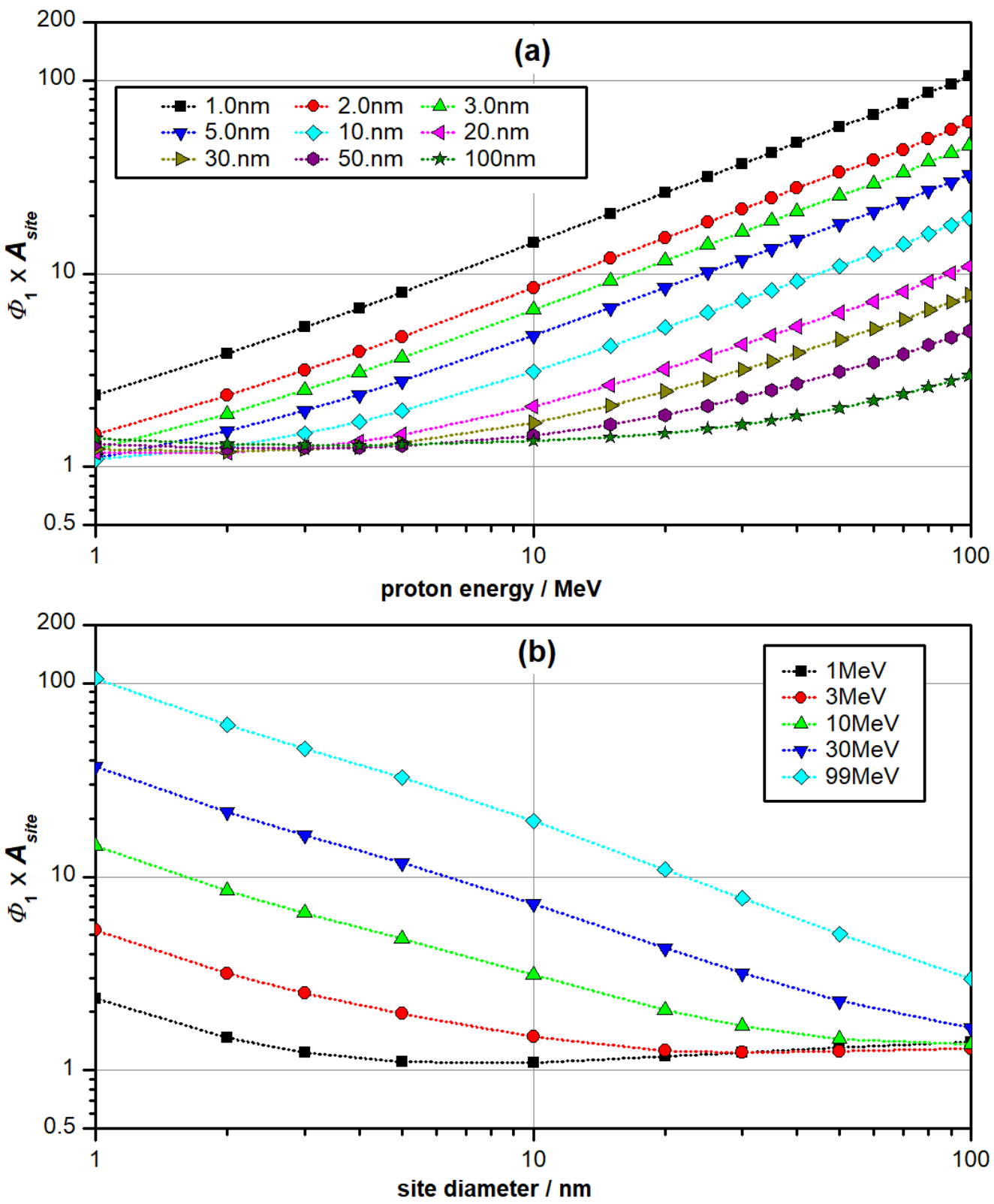

**Supplementary Fig. S11**: Dependence of the product between the fluence producing an event frequency of unity and the cross-section of the site (a) on proton energy for different site sizes (see legend) and (b) on site size for different proton energies (see legend). The data pertain to simulations performed with Geant4-DNA option 6.